\documentclass[journal]{IEEEtranTIE}

\usepackage{booktabs}
\usepackage{colortbl}
\usepackage{amssymb}
\usepackage{algorithmic}
\usepackage{amsmath} 
\usepackage{amsfonts}
\DeclareMathOperator*{\argmin}{argmin} 
\usepackage[linesnumbered,ruled,lined]{algorithm2e}
\usepackage{courier} 
\ifCLASSINFOpdf
\usepackage[pdftex]{graphicx}
\else
\usepackage[dvips]{graphicx}
\fi
\usepackage{epstopdf}
\usepackage{array}
\ifCLASSOPTIONcompsoc
\usepackage[caption=false,font=normalsize,labelfont=sf,textfont=sf]{subfig}
\else
\usepackage[caption=false,font=footnotesize]{subfig}
\fi
\usepackage{color}
\usepackage{xcolor}
\usepackage{textcomp}
\usepackage[normalem]{ulem}
\usepackage{cite}
\usepackage{multibib} 
\usepackage{nomencl}
\makenomenclature
\usepackage{etoolbox}
\renewcommand\nomgroup[1]{%
	\item[\bfseries
	\ifstrequal{#1}{A}{Indices and Sets}{%
		\ifstrequal{#1}{B}{Continuous Decision Variables}{%
			\ifstrequal{#1}{C}{Discrete Decision Variables}{       \ifstrequal{#1}{D}{Parameters}{
			        \ifstrequal{#1}{E}{Dynamic System Variables and Parameters}} }}} ]}

\usepackage{hyperref}
\hypersetup{
	colorlinks=true,
	linkcolor=blue,
	filecolor=magenta,      
	urlcolor=cyan,}
\IEEEoverridecommandlockouts

\begin{document}
\bstctlcite{BSTcontrol}
	
\title{Hybrid Imitation Learning for Real-Time Service Restoration in Resilient Distribution Systems}

\author{Yichen~Zhang,~\IEEEmembership{Member,~IEEE,}
	~Feng~Qiu,~\IEEEmembership{Senior Member,~IEEE,}
	~Tianqi~Hong,~\IEEEmembership{Member,~IEEE,}
	~Zhaoyu~Wang,~\IEEEmembership{Senior Member,~IEEE,}
	~Fangxing~(Fran)~Li,~\IEEEmembership{Fellow,~IEEE}
	\thanks{
	This work was supported by the Advanced Grid Modeling program of U.S. Department of Energy.
	
	Y. Zhang, F. Qiu, T. Hong are with the Energy System Division, Argonne National Laboratory, Lemont, IL 60439 USA (email: yichen.zhang@anl.gov, fqiu@anl.gov, thong@anl.gov). 
	
	Z. Wang is with ECpE Dept., Iowa State University, Ames, IA 50011 USA (email: wzy@iastate.edu).
	
	F. Li is with Department of EECS, University of Tennessee, Knoxville, TN 37996 USA (email: fli6@utk.edu).
	
	The source code of the implementation will be available at \url{https://github.com/ANL-CEEESA/IntelliHealer}.
}}

\markboth{This paper has been accepted by IEEE Transactions on Industrial Informatics (DOI: 10.1109/TII.2021.3078110)}%
{Shell \MakeLowercase{\textit{et al.}}: Bare Demo of IEEEtran.cls for IEEE Journals}

\maketitle

\begin{abstract}
Self-healing capability is a critical factor for a resilient distribution system, which requires intelligent agents to automatically perform service restoration online, including network reconfiguration and reactive power dispatch. The paper proposes the imitation learning framework for training such an agent, where the agent will interact with an expert built based on the mixed-integer program to learn its optimal policy, and therefore significantly improve the training efficiency compared with exploration-dominant reinforcement learning methods. This significantly improved training efficiency makes the training problem under $N-k$ scenarios tractable. A hybrid policy network is proposed to handle tie-line operations and reactive power dispatch simultaneously to further improve the restoration performance.  The 33-bus and 119-bus systems with $N-k$ disturbances are employed to conduct the training. The results indicate that the proposed method outperforms traditional reinforcement learning algorithms such as the deep-Q network.
\end{abstract}

%

\mbox{}
\nomenclature[A,01]{$t$, $\mathcal{T}$, $T$}{index, index set, number of steps}
\nomenclature[A,02]{$h$, $\mathcal{V}_{\text{P}}$, $N_{\text{P}}$}{index, index set, number of point of common coupling}
\nomenclature[A,03]{$i/j$, $\mathcal{V}_{\text{B}}$, $N_{\text{B}}$}{index, index set, number of buses}
\nomenclature[A,04]{$k$, $\mathcal{V}_{\text{SC}}$, $N_{\text{SC}}$}{index, index set, number of shunt capacitors}
\nomenclature[A,06]{$l$, $\mathcal{E}_{\text{L}}$, $N_{\text{L}}$}{index, index set, number index of lines}
\nomenclature[A,07]{$m$, $\mathcal{N}_{\text{S}}$, $N_{\text{S}}$}{index, index set, number (if countable) of states}
\nomenclature[A,08]{$n$, $\mathcal{N}_{\text{A}}$, $N_{\text{A}}$}{index, index set, number (if countable) of actions}

\nomenclature[B,01]{$P_{h,t}^{\text{PCC}}$}{active power injection at point of common coupling $h$ during step $t$}
\nomenclature[B,02]{$Q_{h,t}^{\text{PCC}}$}{reactive power injection at point of common coupling $h$ during step $t$}
\nomenclature[B,03]{$V_{i,t}$}{voltage of bus $i$ during step $t$}
\nomenclature[B,04]{$Q_{k,t}^{\text{SC}}$}{reactive power output of shunt capacitor $k$ during step $t$}
\nomenclature[B,05]{$P_{l,t}$, $Q_{l,t}$}{active, reactive power flow on line $l$ during step $t$}
\nomenclature[B,06]{$\Delta^{\text{SC}}_{k, t}$}{incremental change of shunt $k$ from $t-1$ to $t$}

\nomenclature[C,01]{$u_{l,t}^{\text{L}}$}{status of line $l$ during step $t$: 1 closed and 0 otherwise}
\nomenclature[C,02]{$a_{l,t}^{\text{T}}$}{action decision of tie-line $l$ during step $t$: 1 to be closed and 0 otherwise}
\nomenclature[C,03]{$u_{k,t}^{\text{SC}}$}{status of shunt capacitor $k$ during step $t$: 1 active and 0 otherwise}
\nomenclature[C,04]{$u_{i,t}^{\text{D}}$}{connection status of demand at bus $i$ during step $t$: 1 connected and 0 otherwise}
\nomenclature[C,05]{$u_{i,j,t}^{\text{R}}$}{indication if bus $i$ is the parent bus of $j$: 1 true and 0 false}

\nomenclature[D,01]{$P_{i}^{\text{D}}$, $Q_{i}^{\text{D}}$}{active, reactive power demand at bus $i$}
\nomenclature[D,02]{$\underline{P}_{l}$, $\overline{P}_{l}$}{min, max active power flow of line $l$}
\nomenclature[D,03]{$\underline{Q}_{l}$, $\overline{Q}_{l}$}{min, max reactive power flow of line $l$}
\nomenclature[D,04]{$\underline{Q}^{\text{SC}}_{k}$, $\overline{Q}^{\text{SC}}_{k}$}{min, max reactive power output of shunt capacitor $k$}
\nomenclature[D,05]{$\epsilon$}{allowable voltage deviation from nominal value}
\printnomenclature[0.7in]

\section{Introduction}\label{sec_intro}
Natural disasters can cause random line damages in distribution systems. The distribution system restoration (DSR) is one of the most critical factors to ensure power grid resilience.
The objective of DSR is to search for alternative paths to re-energize the loads in out-of-service areas through a series of switching operations. Typical distribution systems have normally closed sectionalizing switches and normally open tie switches. When a fault is identified, the restoration plan will use tie switches to reconfigure the network so that the disrupted customers can be connected to available feeders \cite{Zidan2017}.

Nowadays, there is an increasing demand to \emph{automatize} the decision-making process of network reconfigurations and DSR, or so-called self-healing. The self-healing capability is considered as one of the most critical factors for a resilient distribution system to reduce the customers minute interruption (CMI) as well as other associated indices like system average interruption duration index (SAIDI), avoiding high interruption cost. For example, the self-healing technology IntelliTeam\footnote{\rmfamily \url{https://www.sandc.com/en/solutions/self-healing-grids/}} developed by S\&C Electric Company will result in a zero CMI and zero SAIDI. While manual restoration usually results in a 90000-minute CMI and 43-minute SAIDI. Under extreme events, the outage will occur more frequently, and the self-healing strategy will have more significant performance in reducing the interruption cost and the overall decision-making complexity. The thrust of the DSR automation is the intelligent agent and the built-in policy mapping from different faulty scenarios to corresponding optimal restorative actions. The methods for building such policies can be categorized into two major types: \emph{predefined} or \emph{reactive}.

The reactive policy requires the agent to solve DSR online once the faulty condition is received. An overview can be found in \cite{Shen2020a}. Currently, most methods rely on mathematical programming (MP). The graph theory was used to search for alternative topology after a fault in \cite{Li2014}. Two novel MP formulations were proposed in \cite{Wang2015b} and \cite{Chen2016a}, respectively, to sectionalize the distribution system into self-sustained microgrids during blackouts. A two-stage heuristic solution was proposed in \cite{Gao2016} to optimize the path from a microgrid source to the critical loads. Multiple distributed generators were incorporated and optimized with a similar objective in \cite{Xu2018a}. Ref. \cite{Shariatzadeh2017} employed the particle swarm optimization (PSO) to solve the shipboard reconfiguration optimization problem. Multi-step optimization formulation were used for restoration in \cite{Wang2017c} and \cite{Chen2018}. In \cite{Arif2018}, network operation and repairing crew dispatch were co-optimized. A relaxed AC power flow formulation was proposed for unbalanced system restoration in \cite{RoofegariNejad2019}. A multi-step reconfiguration model with distributed generators (DG) start-up sequences \cite{Sekhavatmanesh2020}. Distributed optimization with a mixed-integer second-order cone programming problem was formulated in \cite{Shen2020}. The optimization-based multi-agent framework was employed to form self-sustained islands in \cite{Sharma2018}. Ref. \cite{Shirazi2019} proposed three different types of agents that solve a multi-objective optimization to facilitate self-healing. Under similar scopes, a convex optimal power flow model and DGs were considered in \cite{Sekhavatmanesh2019} and \cite{Li2020a}, respectively. However, these technologies need devices to have sophisticated computation architectures. Furthermore, the solution time may not be able to meet the real-time requirement. 

The predefined strategy heavily relies on reinforcement learning (RL) framework to train the policy. The RL framework has been extensively applied into various power system operation problems, including frequency control \cite{Chen2020}, voltage control \cite{Duan2020}, energy management \cite{Du2020}, economic dispatch \cite{Dai2020}, distribution system operation cost reduction via reconfiguration \cite{Gao2020}. While, recent work regarding RL for restoration is limited.
Ref. \cite{Perez-Guerrero2008} employed the dynamic programming algorithm to compute the exact value function, which is intractable for high dimensional problems. In Ref. \cite{Wang2020a}, the value function was estimated using the approximate dynamic programming algorithm.  Both algorithms, however, require the knowledge of the state transition probabilities, which are difficult to know in advance. The temporal difference learning methods, such as Q-learning, estimate the empirical state transition probabilities from observations. In Refs. \cite{Ye2011} and \cite{Das2013}, the Q-learning algorithm with the $\epsilon$-greed policy was employed to perform offline training such that the agent can reconfigure the network online. Ref. \cite{Ghorbani2016} proposed a mixed online and offline strategy, in which the online restoration plan either from the agent or an MP was adopted based on certain confident metrics. While in offline mode, the agent was also trained using the Q-learning algorithm. Despite the innovations, the aforementioned works have not considered random $N-k$ line outages. This disturbance randomness hampers the application of exploration-dominant algorithms like traditional RL, which is known to converge slowly due to the exploration and exploitation dilemma \cite{sutton2018reinforcement}. In other words, these works rely on random exploration strategies, such as $\epsilon$-greed, to locally improve a policy \cite{cheng2020efficient}. With additional disturbance randomness, the number of interactions required to learn a policy is enormous, leading to a prohibitive cost. Such a capability limitation on handling disturbance randomness significantly impedes the deployment in real-world scenarios.

In a nutshell, the major gap of current research for DSR automation can be concluded as
\begin{itemize}
	\item Reactive strategies such as MP-based methods need sophisticated computational architectures and have overrun risk in real-time execution.
	\item Predefined strategies such as RL-based methods have not considered random $N-k$ line outages, which jeopardizes the self-healing capability. The underline reason is that traditional RL is not capable of training the policy efficiently under random disturbances.
\end{itemize}
To overcome this limitation, the paper employs the predefined strategy and proposes the imitation learning (IL) framework for training the restoration agent. 
The advantages of IL methods are the significantly higher training efficiency since it leverages prior knowledge about a problem in terms of expert demonstrations and trains the agents to mimic these demonstrations. With the proposed method, random $N-k$ disturbances are tractable and considered in this paper to enhance the self-healing capability under various conditions. 
Its fundamental form consists of training a policy to predict the expert’s actions from states in the demonstration data using supervised learning. Here, we leverage well-studied MP-based restoration as the expert. In addition, reconfigured networks may exhibit longer lines and low voltages. Thus, tie-line operations and reactive power dispatch are considered simultaneously to restore more loads. The contribution of this paper is concluded as follows
\begin{itemize}
	\item proposing the IL framework to improve training efficiency and reduce the number of agent-environment interactions required to train a policy. We show that the IL algorithms significantly outperform the deep Q-learning under random $N-k$ contingencies.
	\item strategically designing MP-based experts and environments with tailored formulations for IL algorithms
	\item developing a hybrid policy structure and training algorithms to accommodate the mixed discrete and continuous action space
\end{itemize}
Concisely, this paper proposes to use the new IL paradigm for training DSR policy that is capable of handling training complexity under random disturbances. The proposed solution can successfully handle the central technical requirements of self-healing capability, that is, automatic and optimal.
It is worth mentioning that the IL framework acts as a bridge between RL-based techniques and MP-based methods and a way to leverage well-studied MP-based decision-making systems for RL-based automation.
The source code of the implementation will be available at \url{https://github.com/ANL-CEEESA/IntelliHealer}.

The remainder of this paper is organized as follows. Section \ref{sec_prob} will frame the DSR problem as the Markov decision process (MDP). Section \ref{sec_il} will introduce the IL problem and algorithms. Section \ref{sec_expert} will introduce the MP-based experts and environments that IL algorithms are interacting with. Section \ref{sec_case} will illustrate the case study, followed by the conclusion in \ref{sec_con}.

\section{Problem Statement}\label{sec_prob}
Let the distribution system be denoted as a graph $\mathcal{G}=(\mathcal{V}_{\text{B}},\mathcal{E}_{\text{L}})$, where $\mathcal{V}_{\text{B}}$ denotes all buses (vertices) and $\mathcal{E}_{\text{L}}$ denotes all lines (edges). The bus set is categorized into substation buses $\mathcal{V}_{\text{B,S}}$ and non-substation buses $\mathcal{V}_{\text{B,NS}}$. The line set is categorized into non-switchable line set $\mathcal{E}_{\text{L,NS}}$ and tie-line set $\mathcal{E}_{\text{L,T}}$. The non-switchable lines can not be actively controlled unless tripped due to external disturbances. The status of tie-lines can be controlled through tie-switches to adjust the network configuration. 

Assume a $N_{\text{L,NS}}-k$ contingency scenario indicating that $k$ lines from the set $\mathcal{E}_{\text{L,NS}}$ are tripped. Without loss of generality, we uniformly sample these $k$ lines from $\mathcal{E}_{\text{L,NS}}$ in each scenario (or episode\footnote{\rmfamily The terms \emph{scenario} and \emph{episode} are regarded the same in this paper and will be used interchangeably.}). Let $\mathcal{E}_{\text{L,NS}}^{\text{F}}$ be the set of faulty lines and $\mathcal{E}_{\text{L,NS}}^{\text{NF}}$ be the set of non-faulty lines. The goal for a well-trained agent is to control the tie-lines and shunt capacitors to optimally restore interrupted customers given post-fault line status.

To account for the time-dependent process \cite{Sekhavatmanesh2020}, such as the saturating delays of tie-switches and shunt capacitors, as well as reducing transients, we consider a multi-step restoration. In each step, \emph{only one} tie-line is allowed to operate. In addition, closed tie-lines are not allowed to open again. Meanwhile, all shunt capacitors can be dispatched at each step. Based on state-of-the-art industrial self-healing productions, such as IntelliTeam from S\&C Electric Company and Distribution Feeder Automation (SDFA) from Siemens Industry Inc., the time interval between each step is in the time scale of seconds. The specific value can vary between systems due to different topology, power sources and devices. Naturally, the step number is set to be equal to the number of tie-lines $N_{\text{L,T}}$ in the system, that is, $\mathcal{T} = [0,1,\cdots,N_{\text{L,T}}]$, where $0$ stands for the initial step. A positive integer will be used to label each tie-line action as shown in Eq. (\ref{eq_action_space_1}). In many scenarios, \emph{not} all tie-lines are involved. For a step $t\in\mathcal{T}$ where no tie-line action is needed, the action label will be zero. During the restoration process, the network radiality must be maintained, and the tie-line operations that violate the radiality constraint will be denied. We formalize the above setting using the episodic finite Markov decision process (EF-MDP) \cite{sutton2018reinforcement}. An EF-MPD $\mathcal{M}$ can be described by a six-tuple $\mathcal{M}=\textless\mathcal{S}, \mathcal{A}, \mathcal{D}, p(s^{\prime}|s,a), r(s,a), T\textgreater$, where $\mathcal{S}$ denotes the state space, $\mathcal{A}$ denotes the action space, $\mathcal{D}$ denotes the disturbance space, $p(s^{\prime}|s,a)$ denotes the state transition probability, $r$ denotes the real-valued reward function, $T$ denotes the number of steps in each episode, and $s^{\prime}, s\in\mathcal{S}$, $a\in \mathcal{A}$. The action space is hybrid, consisting of a discrete action space $\mathcal{A}_{\text{T}}$ for tie-line operations and a continuous action space $\mathcal{A}_{\text{C}}$ where
\begin{align}
& \mathcal{A}_{\text{T}}=[0,1,\cdots,N_{\text{L,T}}]\label{eq_action_space_1}\\
& \mathcal{A}_{\text{C}}=[\underline{Q}^{\text{C}}_{1}, \overline{Q}_{1}^{\text{C}}]\cup\cdots\cup[\underline{Q}^{\text{C}}_{N_{\text{C}}}, \overline{Q}^{\text{C}}_{N_{\text{C}}}]\label{eq_action_space_2}
\end{align}
A trajectory can be denoted as 
\begin{align}
\tau=(s_{0}(d), a_{1}, s_{1}, a_{2}, s_{2},\cdots,a_{T}, s_{T})
\end{align}
where $s_{0}(d)$, or $s_{0}$ for short, is the initial faulty condition due to disturbance $d\in\mathcal{D}$. For actions that violate the radiality constraint, the corresponding transition probability will be zero and one otherwise.


\section{Deep Imitation Learning}\label{sec_il}
\subsection{Imitation Learning Problem}
The IL training process aims to search for a policy $\pi(a|s)$ (a conditional distribution of action $a\in\mathcal{A}$ given state $s\in\mathcal{S}$) from the class of policies $\Pi$ to mimic the expert policy $\pi^{*}$ \cite{ross2010efficient}. The expert policy is assumed to be deterministic. Without loss of generality, consider a countable state space $\mathcal{S}=[s^{1},s^{2},\cdots,s^{N_{\text{S}}}]$ with $N_{\text{S}}$ number states. Let $\rho_{0}$ denote the initial distribution of states and $\rho_{0}(s^{m})$ denote the probability of state $s^{m}$. Let $\rho_{t}^{\pi}$ denote the distribution of states at time $t$ if the agent executes the policy $\pi$ from step 1 to $t-1$. The law of $\rho_{t}^{\pi}$ can be computed recursively as follows \cite{ross2010efficient}
\begin{align}
\rho_{t}^{\pi}(s^{m}_{t})=\sum_{s_{t-1}\in\mathcal{S}}\rho_{t-1}(s_{t-1})\sum_{a_{t}\in\mathcal{A}}\pi(a_{t}|s_{t-1})p(s^{m}_{t}|s_{t-1},a_{t})
\end{align}
Then, the average distribution of states is defined as $\bar{\rho}^{\pi}(s)=\sum_{t=1}^{T}\rho_{t-1}^{\pi}(s)/T$, which represents the state visitation frequency over $T$ time steps if policy $\pi$ is employed \cite{sun2019towards}.

The 0-1 loss of executing action $a$ in state $s$ with respect to (w.r.t) the expert policy $\pi^{*}$ is denoted as follows \cite{ross2011reduction}
\begin{align}
e(s,a) = I(a \neq \pi^{*}(s))
\end{align}
where $I(\bullet)$ is the indicator function. Consider an action $a_{t}$. If this action is different from the action provided by the optimal policy $a^{*}=\pi^{*}(s)$, that is, $a_{t}\neq a^{*}$, then the loss value equals one. Otherwise, it equals zero. Intuitively, if an agent is able to act identically as the optimal policy, the loss will be zero. The expected 0-1 loss of policy $\pi$ in state $s$ reads as follows
\begin{align}
e_{\pi}(s) = \mathbb{E}_{a\sim\pi{_{s}}}[e(s,a)]
\end{align}
The expected $T$-step loss w.r.t $\pi$ is
\begin{align}
L(\pi) = \mathbb{E}_{s\sim \rho^{\pi}}[e_{\pi}(s)]
\end{align}
The goal is to find a policy $\bar{\pi}$ that minimize the expected $T$-step loss $L(\pi)$, that is,
\begin{align}
\label{eq_il_objective}
\bar{\pi} = \argmin_{\pi\in\Pi}L(\pi) = \argmin_{\pi\in\Pi}\mathbb{E}_{s\sim \rho^{\pi}}[e_{\pi}(s)]
\end{align}
Note that this objective function is non-convex due to the dependence between the objective parameter $\rho^{\pi}$ and the decision space $\Pi$.

\subsection{Imitation Learning Algorithm}
The most effective form of imitation learning is behavior cloning (BC). In the BC algorithm summarized, trajectories are collected under the expert's policy $\pi^{*}$, and the IL problem renders to a supervised learning problem, where the states are the features, and the actions are the labels. The objective of BC reads as follows 
\begin{align}
\label{eq_bc_objective}
\bar{\pi} = \argmin_{\pi\in\Pi}\mathbb{E}_{s\sim \rho^{\pi^{*}}}[e_{\pi}(s)]
\end{align}
which disassociates the dependency between the objective parameter and the decision space \cite{ross2011reduction}. 

\begin{figure}[h]
	\centering
	\includegraphics[scale=0.25]{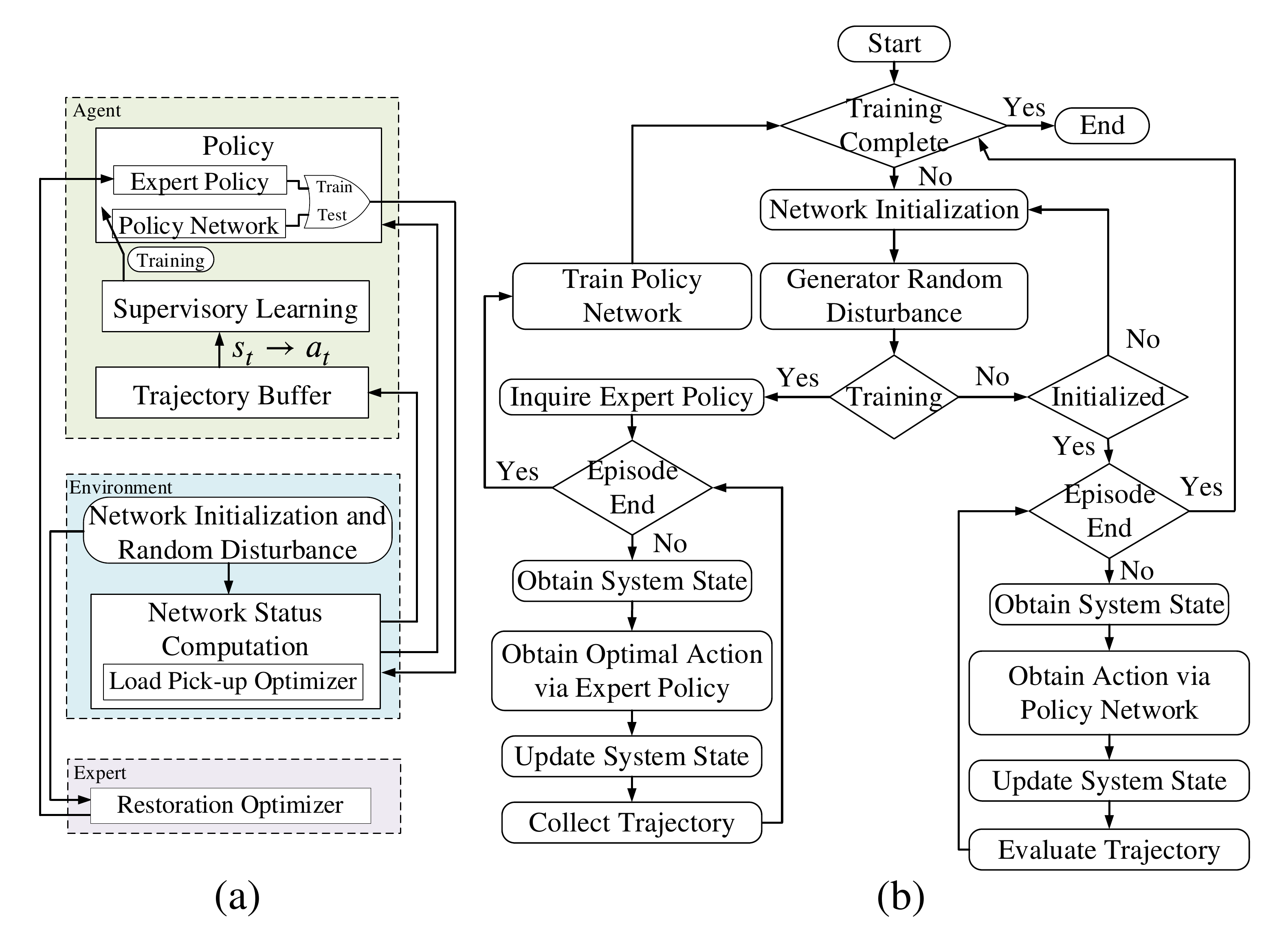}
	\caption{IL framework and algorithm for training DSR agent. (a) IL paradigm. (b) IL algorithm flowchart.}
	\label{fig_IL_flowchart}
\end{figure}

A framework paradigm and the basic algorithm flowchart for IL-based DSR agent are illustrated in Fig. \ref{fig_IL_flowchart} (a) and (b), respectively. Three modules, including agent, environment, and expert, will be built, as shown in Fig. \ref{fig_IL_flowchart} (a). In each training step, the agent will use the expert policy to explore the environment to obtain the optimal trajectory for training. Then, testing will be conducted under a new disturbance with the trained policy network, as illustrated in Fig. \ref{fig_IL_flowchart} (b).

The BC algorithm is described in Algorithm \ref{algo_bc}. Several major functions are explained as follows.
\begin{itemize}
	\item $\texttt{Expert}$: Since we are addressing a multi-period scheduling problem, it is difficult to directly obtain an expert mapping $\pi^{*}$. Therefore, a mixed-integer program (MIP) is employed to obtain the optimal actions. This MIP is specified as an expert solver $\texttt{Expert}(s_{t-1},[t,\cdots,T])$, which takes the initial state at $t-1$ and the scheduling interval $[t,\cdots,T]$, and return the optimal actions $a_{t},\cdots,a_{T}$. The detailed MIP formulation is given in Section \ref{sec_expert}.
	\item $\texttt{Act}$: The DSR environment interacts with the policy through $\texttt{Act}$. Given a disturbance $d$, total step $T$, and the policy (either the mapping or expert solver), $\texttt{Act}$ returns a $T$-step trajectory. More details are described in Algorithm \ref{algo_act}.
	\item $\texttt{Eval}$: $\texttt{Eval}$ compares the learned policy-induced trajectory with the optimal one that is provided by the MP-based method, or the expert, and calculates the ratio $r$ between restored total energy under the learned policy and the optimal restored total energy. The ratio is defined as the performance score of the learned policy in each iteration.
\end{itemize}

\begin{algorithm}
	\SetKwData{Left}{left}\SetKwData{This}{this}\SetKwData{Up}{up}
	\SetKwFunction{Sample}{Sample}\SetKwFunction{Expert}{Expert}
	\SetKwFunction{Act}{Act}\SetKwFunction{ActMix}{ActMix}
	\SetKwFunction{Env}{Env}\SetKwFunction{Reset}{Reset}\SetKwFunction{Step}{Step}
	\SetKwFunction{Train}{Train}\SetKwFunction{Eval}{Eval}
	\SetKwInOut{Input}{input}\SetKwInOut{Output}{output}
	
	\Input{expert solver $\Expert$, deep neural net policy $\bar{\pi}$, neural network training function $\Train(\cdot,\cdot,\cdot)$, environment interaction $\Act(\cdot,\cdot,\cdot)$, disturbance set $\mathcal{D}$, stochastic sampling function $\Sample(\cdot)$, policy evaluation function $\Eval(\cdot,\cdot)$}
	
	$X \leftarrow \emptyset$ \tcp*[f]{initialize the input}
	
	$Y \leftarrow \emptyset$ \tcp*[f]{initialize the label}
	
	$P \leftarrow \emptyset$ \tcp*[f]{initialize the performance}
	
	$\bar{\pi}^{1}\in\Pi$ \tcp*[f]{initialize the policy}
	
	\For{$i \leftarrow 1$ \KwTo $N$}{
		
		$d\leftarrow\Sample(\mathcal{D})$
		
		$(s_{0},a_{1},s_{1},\cdots,a_{T},s_{T})\leftarrow\Act(d, T, \Expert)$
		
		$X \leftarrow X\cup (s_{0},\cdots,s_{T-1})$
		
		$Y \leftarrow Y\cup (a_{1},\cdots,a_{T})$
			
		$ \bar{\pi}^{i+1} \leftarrow \Train(X,Y,\bar{\pi}^{i})$\label{algo_1_train}
		
		$d\leftarrow\Sample(\mathcal{D})$
		
		$r \leftarrow \Eval(\Act(d, T, \Expert), \Act(d,T,\bar{\pi}^{i+1}))$
		
		$P \leftarrow P\cup (d,r)$

	}
	\Output{Trained deep neural net $\bar{\pi}$, performance scores $P$}
	\caption{Behavior cloning (BC)}\label{algo_bc}
\end{algorithm}

Algorithm \ref{algo_act} runs either the learned policy or the expert solver on the DSR environment $\texttt{Env}$ to obtain the trajectory. The DSR environment $\texttt{Env}$ is built on the standard Open-AI Gym environment template \cite{1606.01540}. There are two major functions: $\texttt{Env.Reset}$ and $\texttt{Env.Step}$. 
\begin{itemize}
	\item $\texttt{Env.Reset}$ generates a certain number of line outages, computes the initial system status under the line outages using Eq. (\ref{eq_env_reset}), and updates the system states and actions.
	\item $\texttt{Env.Step}$ receives a tie-line action, solve the MIP program Eq. (\ref{eq_env_step}) to obtain the new system state, and updates the system states and actions. If this action violates the radiality and other technical constraints and results in infeasibility, the action will be denied and thus will not be updated. The reward will be computed based on restored loads. Note that the rewards are specifically calculated for RL as IL does not need rewards.
\end{itemize}
  
\begin{algorithm}
	\SetKwData{Left}{left}\SetKwData{This}{this}\SetKwData{Up}{up}
	\SetKwFunction{Sample}{Sample}\SetKwFunction{Expert}{Expert}
	\SetKwFunction{Act}{Act}\SetKwFunction{ActMix}{ActMix}
	\SetKwFunction{Env}{Env}\SetKwFunction{Reset}{Reset}\SetKwFunction{Step}{Step}
	\SetKwFunction{Train}{Train}\SetKwFunction{Eval}{Eval}
	\SetKwInOut{Input}{input}\SetKwInOut{Output}{output}
	
	\Input{disturbance $d$, time step $T$, policy function or expert solver $f$, DSR environment $\Env$}

	$s_{0} \leftarrow \Env.\Reset(d)$
	
	\If{$f==\Expert$} 
	{\tcc{run Env under the expert policy}
		$(a_{1},\cdots,a_{T}) \leftarrow \Expert(s_{0},[1,\cdots,T])$
			
		\For{$t \leftarrow 1$ \KwTo $T$}
		{
			$s_{t}\leftarrow \Env.\Step(a_{t})$	
		}
	}

	\If{$f==\pi$}
	{\tcc{run Env under learned policy}
		\For{$t \leftarrow 1$ \KwTo $T$}
		{
			$a_{t} \leftarrow \pi(s_{t-1})$
			
			$s_{t}\leftarrow \Env.\Step(a_{t})$
		}
	}

	\Output{$T$-step trajectory $(s_{0},a_{1},s_{1},\cdots,a_{T},s_{T})$}
	\caption{Environment interaction \texttt{Act}}\label{algo_act}
\end{algorithm}

\subsection{Hybrid Policy}
\begin{figure}[h]
	\centering
	\includegraphics[scale=0.3]{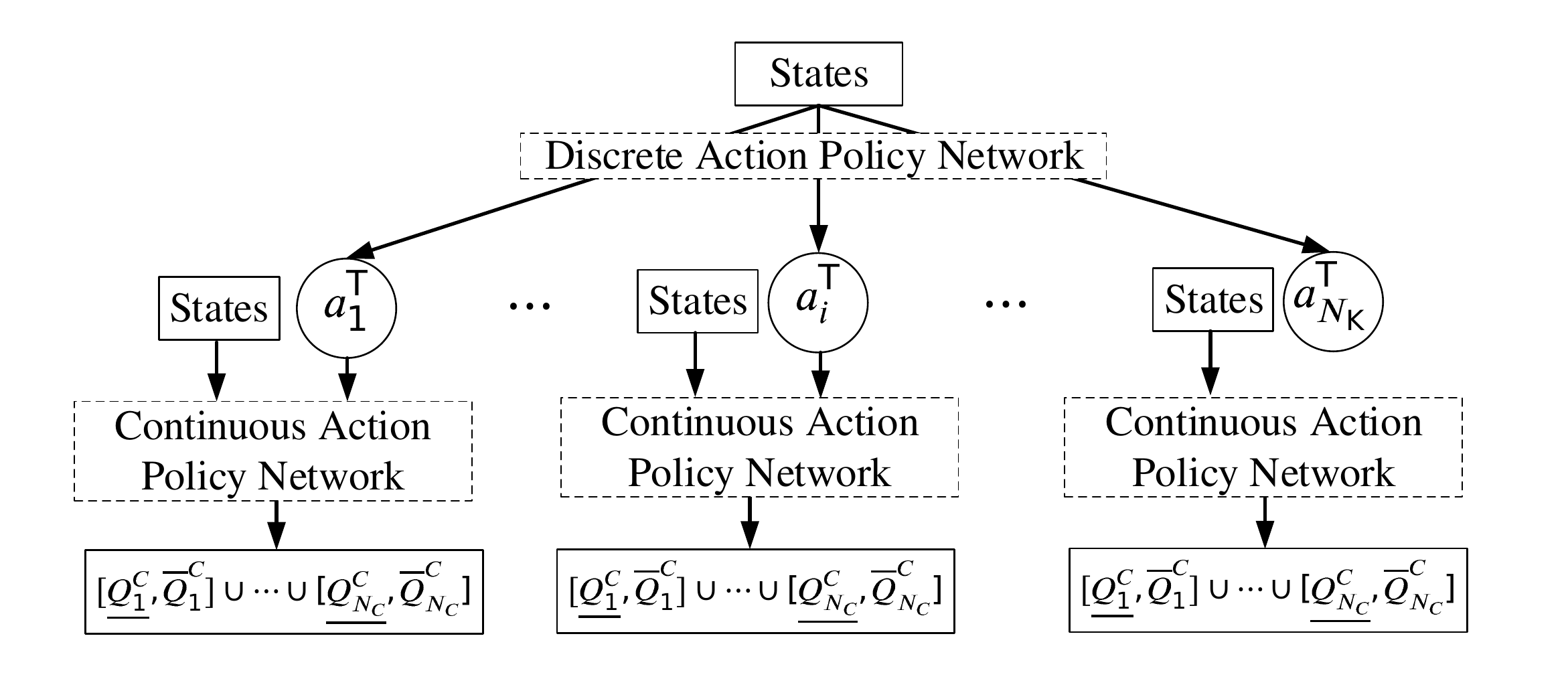}
	\caption{Discrete-continuous hybrid policy network.}
	\label{fig_hybrid_network}
\end{figure}

The training in Algorithm \ref{algo_bc} Line \ref{algo_1_train} is a multi-class classification problem, which is not able to handle continuous action spaces. Thus, Algorithm \ref{algo_bc} can only be used for automatic tie-line operators. To simultaneously coordinate tie-line operations and reactive power dispatch, we propose a hybrid policy network, as shown in Fig. \ref{fig_hybrid_network}. The action spaces of the hybrid neural network are mixed continuous and discrete. At the higher level, there is a single neural network to predict the optimal tie-line actions given measured states. Each tie-line action is associated with a neural network for reactive power dispatch. The dispatch ranges associated with individual tie-lines can be a subset or entire continuous action spaces. Considering the fact that under each tie-line operation, the system may admit a different power flow pattern, we attach the entire dispatch spaces in each tie-line action. It is also worth mentioning that the states for predicting discrete and continuous actions can be different.

The training process for the hybrid policy network is described in Algorithm \ref{algo_hybrid_bc}. The additional effort from Algorithm \ref{algo_bc} is that we will train reactive power dispatchers under each tie-line action. To do this, we first initialize the dispatcher training dataset as shown in Line \ref{algo_3_int}. In each episode, we group the dispatch commands from the expert \texttt{hExp} based on the tie-line actions as shown in Lines \ref{algo_3_x} and \ref{algo_3_y}. The final step in each episode is to train the tie-line operation policy network and reactive power dispatch policy network, respectively, as shown in Lines \ref{algo_3_clf} and \ref{algo_3_reg}. The hybrid behavior cloning algorithm will interact with the environment that includes both tie-line and reactive power dispatch, which is described in Algorithm \ref{algo_hybrid_act}. Algorithm \ref{algo_hybrid_act} is similar to Algorithm \ref{algo_act} except that the hybrid actions are generated using the hybrid policy as shown in Lines \ref{algo_4_act1} and \ref{algo_4_act2}, and the DSR environment has hybrid actions. The MIP formulation of \texttt{hEnv} will be introduced in Section \ref{sec_expert}.
\begin{algorithm}
	\SetKwData{Left}{left}\SetKwData{This}{this}\SetKwData{Up}{up}
	\SetKwFunction{Sample}{Sample}\SetKwFunction{hExp}{hExp}
	\SetKwFunction{Act}{Act}\SetKwFunction{hAct}{hAct}
	\SetKwFunction{Env}{Env}\SetKwFunction{Reset}{Reset}\SetKwFunction{Step}{Step}
	\SetKwFunction{TrainClf}{TrainClf}\SetKwFunction{TrainReg}{TrainReg}\SetKwFunction{Eval}{Eval}
	\SetKwInOut{Input}{input}\SetKwInOut{Output}{output}
	
	\Input{hybrid action expert solver $\hExp$, tie-line operation policy $\bar{\pi}$, reactive power dispatch policy under tie-line action $k$ $\bar{\pi}_{k}$, tie-line operation policy network training function $\TrainClf(\cdot,\cdot,\cdot)$, reactive power dispatch policy network training function $\TrainReg(\cdot,\cdot,\cdot)$, hybrid action environment interaction $\hAct(\cdot,\cdot,\cdot)$, disturbance set $\mathcal{D}$, stochastic sampling function $\Sample(\cdot)$, policy evaluation function $\Eval(\cdot,\cdot)$}
	
	$X \leftarrow \emptyset$, $Y \leftarrow \emptyset$
	
	$X_{k} \leftarrow \emptyset$, $Y_{k} \leftarrow \emptyset$\label{algo_3_int}
	
	$P \leftarrow \emptyset$
	
	$\bar{\pi}^{1}\in\Pi$, $\bar{\pi}^{1}_{k}\in\Pi$
	
	\For{$i \leftarrow 1$ \KwTo $N$}{
		
		$d\leftarrow\Sample(\mathcal{D})$
		
		$(s_{0},a^{\text{D}}_{1},a^{\text{C}}_{1},s_{1},\cdots,a_{T}^{\text{D}},a_{T}^{\text{C}},s_{T})\leftarrow\hAct(d, T, \hExp)$
		
		$X \leftarrow X\cup (s_{0},\cdots,s_{T-1})$
		
		$Y \leftarrow Y\cup (a^{\text{D}}_{1},\cdots,a^{\text{D}}_{T})$
		
		\For{$t \leftarrow 1$ \KwTo $T$}{
			$X_{a^{\text{D}}_{t}} \leftarrow X_{a^{\text{D}}_{t}}\cup s_{t-1}$\label{algo_3_x}
		
			$Y_{a^{\text{D}}_{t}} \leftarrow Y_{a^{\text{D}}_{t}}\cup a^{\text{C}}_{t}$\label{algo_3_y}
		}
	
		$ \bar{\pi}^{i+1} \leftarrow \TrainClf(X,Y,\bar{\pi}^{i})$\label{algo_3_clf}
		
		$ \bar{\pi}^{i+1}_{k} \leftarrow \TrainReg(X_{k},Y_{k},\bar{\pi}^{i}_{k})$\label{algo_3_reg}
		
		$d\leftarrow\Sample(\mathcal{D})$
		
		$r \leftarrow \Eval(\hAct(d, T, \hExp), \hAct(d,T,(\bar{\pi}^{i+1},\bar{\pi}^{i+1}_{k})))$
		
		$P \leftarrow P\cup (d,r)$
		
	}
	\Output{Trained tie-line operator $\bar{\pi}$, trained reactive power dispatcher $\bar{\pi}_{k}$, performance scores $P$}
	\caption{Hybrid behavior cloning (HBC)}\label{algo_hybrid_bc}
\end{algorithm}
\begin{algorithm}
	\SetKwData{Left}{left}\SetKwData{This}{this}\SetKwData{Up}{up}
	\SetKwFunction{Sample}{Sample}\SetKwFunction{hExp}{hExp}
	\SetKwFunction{Act}{Act}\SetKwFunction{ActMix}{ActMix}
	\SetKwFunction{hEnv}{hEnv}\SetKwFunction{Reset}{Reset}\SetKwFunction{Step}{Step}
	\SetKwFunction{Train}{Train}\SetKwFunction{Eval}{Eval}
	\SetKwInOut{Input}{input}\SetKwInOut{Output}{output}
	
	\Input{disturbance $d$, time step $T$, policy function or expert solver $f$, hybrid-action DSR environment $\hEnv$}
	
	$s_{0} \leftarrow \hEnv.\Reset(d)$
	
	\If{$f==\hExp$} 
	{
		$(a_{1}^{\text{D}},a_{1}^{\text{C}},\cdots,a_{T}^{\text{D}},a_{T}^{\text{C}}) \leftarrow \hExp(s_{0},[1,\cdots,T])$
		
		\For{$t \leftarrow 1$ \KwTo $T$}
		{
			$s_{t}\leftarrow \hEnv.\Step((a_{t}^{\text{D}},a_{t}^{\text{C}}))$	
		}
	}
	
	\If{$f==(\pi,\pi_k)$}
	{
		\For{$t \leftarrow 1$ \KwTo $T$}
		{
			$a_{t}^{\text{D}} \leftarrow \pi(s_{t-1})$\label{algo_4_act1}
			
			$a_{t}^{\text{C}} \leftarrow \pi_{a_{t}^{\text{D}}}(s_{t-1})$\label{algo_4_act2}
			
			$s_{t}\leftarrow \hEnv.\Step((a_{t}^{\text{D}},a_{t}^{\text{C}}))$
		}
	}
	
	\Output{$T$-step trajectory $(s_{0},a_{1}^{\text{D}},a_{1}^{\text{C}},s_{1},\cdots,a_{T}^{\text{D}},a_{T}^{\text{C}},s_{T})$}
	\caption{Hybrid action environment interaction \texttt{hAct}}\label{algo_hybrid_act}
\end{algorithm}

\section{Mathematical Programming-Based Expert and Environment}\label{sec_expert}
This section describes the MIP formulation for the experts and environments. We will first introduce generic constraints for the DSR problem. Then, \texttt{Expert}, \texttt{Env.Reset}, \texttt{Env.Step}, \texttt{hExp}, \texttt{hEnv.Reset} and \texttt{hEnv.Step} are introduced in Eqs. (\ref{eq_expert}), (\ref{eq_env_reset}), (\ref{eq_env_step}), (\ref{eq_h_expert}), (\ref{eq_h_env_reset}), (\ref{eq_h_env_step}), respectively.

Let $\mathcal{L}(\cdot,i)$ denote the set of lines for which bus $i$ is the to-bus, and $\mathcal{L}(i,\cdot)$ denote the set of lines for which bus $i$ is the from-bus. Let $\mu(l)$ and $\nu(l)$ map from the index of line $l$ to the index of its from-bus and to-bus, respectively. The nature of radiality guarantees that $\mu(l)$ and $\nu(l)$ are one-to-one mappings. Let $\mathcal{P}$ map from the index of bus $i$ to the substation index. Without loss of generality, we consider one active substation and assume Bus 1 is connected to it. Let $\mathcal{C}$ map from the index of bus $i$ to the shunt capacitor. Let $\mathcal{T}=[t_0, t_1, \cdots, T]$ be the step index and $t\in\mathcal{T}$. 

Following the convention in \cite{Wang2015} and \cite{Arif2018}, linearized Distflow equations are employed to represent power flows and voltages in the network and are described as follows
\begin{align}
\label{eq_con_distflow_1}
\begin{aligned}
\sum_{\forall l\in \mathcal{L}(\cdot,i)}P_{l,t}  &+ \sum_{\forall h\in \mathcal{P}(i)}P^{\text{PCC}}_{h,t}\\
&= \sum_{\forall l\in \mathcal{L}(i,\cdot)}P_{l,t} + u_{i,t}^{\text{D}}P^{\text{D}}_{i,t}\quad\forall i,\forall t\\
\sum_{\forall l\in \mathcal{L}(\cdot,i)}Q_{l,t} & + \sum_{\forall h\in \mathcal{P}(i)}Q^{\text{PCC}}_{h,t} + \sum_{\forall k\in \mathcal{C}(i)}Q^{\text{SC}}_{k,t}\\
&= \sum_{\forall l\in \mathcal{L}(i,\cdot)}Q_{l,t} + u_{i,t}^{\text{D}}Q^{\text{D}}_{i,t}\quad\forall i,\forall t
\end{aligned}
\end{align}
The line flow should respect the limits, which will be enforced to be zero if it is opened
\begin{align}
\label{eq_con_distflow_2}
\begin{aligned}
& u^{\text{L}}_{l,t}\underline{P}_{l}\leq P_{l,t}\leq u^{\text{L}}_{l,t}\overline{P}_{l}\quad\forall l,\forall t\\
& u^{\text{L}}_{l,t}\underline{Q}_{l}\leq Q_{l,t}\leq u^{\text{L}}_{l,t}\overline{Q}_{l}\quad\forall l,\forall t
\end{aligned}
\end{align}
The shunt capacitor should also respect the limits, which will be enforced to be zero if it is opened
\begin{align}
\label{eq_con_shunt}
\begin{aligned}
& u^{\text{SC}}_{k,t}\underline{Q}_{k}^{\text{SC}}\leq Q^{\text{SC}}_{k,t}\leq u^{\text{SC}}_{k,t}\overline{Q}_{k}^{\text{SC}}\quad\forall l,\forall t
\end{aligned}
\end{align}
The linear relation between voltages and line flow needs to be enforced when the line $l$ is closed
\begin{align}
\label{eq_con_distflow_3}
\begin{aligned}
& (u^{\text{L}}_{l,t}-1)M \leq V_{\nu(l),t}-V_{\mu(l),t} + \dfrac{R_{l}P_{l,t}+X_{l}Q_{l,t}}{V_{1}} \quad\forall l,\forall t\\
& (1-u^{\text{L}}_{l,t})M \geq V_{\nu(l),t}-V_{\mu(l),t} + \dfrac{R_{l}P_{l,t}+X_{l}Q_{l,t}}{V_{1}} \quad\forall l,\forall t
\end{aligned}
\end{align}
The voltages should be maintained within permissible ranges
\begin{align}
\label{eq_con_volt}
& 1-\epsilon\leq V_{i,t} \leq 1+\epsilon\quad\forall i,\forall t
\end{align}
The radiality constraints are expressed as follows \cite{Jabr2012}
\begin{align}
\label{eq_con_radial}
\begin{aligned}
& u^{\text{R}}_{\mu(l),\nu(l),t}+u^{\text{R}}_{\nu(l),\mu(l),t}=u^{\text{L}}_{l,t}\quad\forall l,\forall t\\
& u^{\text{R}}_{i,j,t}= 0\quad\forall i,\forall j\in\mathcal{V}_{\text{B,S}},\forall t\\
& \sum_{i\in\mathcal{N}_{\text{B}}}u^{\text{R}}_{i,j,t}\leq 1\quad\forall j,\forall t
\end{aligned}
\end{align}
It is worth mentioning that we may exhibit islands in certain $N-k$ scenarios, where radiality is enforced by Eq. (\ref{eq_con_radial}) only in the main energized branch \cite{Ahmadi2015}. For non-energized islands, loop may exist. But as long as they are energized, Eq. (\ref{eq_con_radial}) becomes effective again. Within all non-switchable lines $\mathcal{E}_{\text{L,NS}}$, the status of faulty lines  $\mathcal{E}_{\text{L,NS}}^{\text{F}}$ is enforced to be zero and the status of non-faulty lines  $\mathcal{E}_{\text{L,NS}}^{\text{NF}}$ is enforced to be one
\begin{align}
\label{eq_con_line}
\begin{aligned}
& u^{\text{L}}_{l,t}=0\quad\forall l\in\mathcal{E}_{\text{L,NS}}^{\text{F}},\forall t\\
& u^{\text{L}}_{l,t}=1\quad\forall l\in\mathcal{E}_{\text{L,NS}}^{\text{NF}},\forall t
\end{aligned}
\end{align}
For a multi-step scenario, the restored loads are not allowed to be disconnected again
\begin{align}
\label{eq_con_load}
\begin{aligned}
& u^{\text{D}}_{i,t}\geq u^{\text{D}}_{i,t-1}\quad\forall i,\forall t\setminus\{t_0\}\\
\end{aligned}
\end{align}
Similarly, closed tie-lines cannot be opened
\begin{align}
\label{eq_con_tieline_1}
\begin{aligned}
& u^{\text{L}}_{l,t}\geq u^{\text{L}}_{l,t-1}\quad\forall l\in\mathcal{E}_{\text{L,T}},\forall t\setminus\{t_0\}\\
\end{aligned}
\end{align}
In addition, only one tie-line can be operated in one step
\begin{align}
\label{eq_con_tieline_2}
\begin{aligned}
& \sum_{l\in\mathcal{N}_{\text{L,T}}}u^{\text{L}}_{l,t}-\sum_{l\in\mathcal{E}_{\text{L,T}}}u^{\text{L}}_{l,t-1}\leq 1\quad\forall t\setminus\{t_0\}\\
\end{aligned}
\end{align}
And all tie-lines are equal to the initial values
\begin{align}
\label{eq_con_tieline_3}
\begin{aligned}
& u^{\text{L}}_{l,t_0}=\hat{u}^{\text{L}}_{l}\quad\forall l\in\mathcal{E}_{\text{L,T}}
\end{aligned}
\end{align}
In some instances, there will be multiple shunt capacitor dispatch solutions for an optimal load restoration, and the shunt dispatch results will jumpy between these solutions in an episode. This will jeopardize a smooth learning process. Therefore, a set of constraints is considered to limit the dispatch frequency
\begin{subequations}
\label{eq_con_sc_switch}
\begin{align}
M(1-z_{k,t}) &\leq Q^{\text{SC}}_{k,t} - Q^{\text{SC}}_{k,t-1}\label{eq_con_sc_switch_1}\\
-M(1-z_{k,t}) &\leq \Delta^{\text{SC}}_{k,t} - (Q^{\text{SC}}_{k,t} - Q^{\text{SC}}_{k,t-1}) \label{eq_con_sc_switch_2}\\
M(1-z_{k,t})  &\geq \Delta^{\text{SC}}_{k,t} - (Q^{\text{SC}}_{k,t} - Q^{\text{SC}}_{k,t-1}) \label{eq_con_sc_switch_3}\\
-Mz_{k,t} &\leq \Delta^{\text{SC}}_{k,t} + (Q^{\text{SC}}_{k,t} - Q^{\text{SC}}_{k,t-1}) \label{eq_con_sc_switch_4}\\
Mz_{k,t} &\geq \Delta^{\text{SC}}_{k,t} + (Q^{\text{SC}}_{k,t} - Q^{\text{SC}}_{k,t-1}) \label{eq_con_sc_switch_5}\\
&\forall k, \forall t \setminus\{t_0\}
\end{align}
\end{subequations}
where we introduce two slack variables: $\Delta^{\text{SC}}_{k,t}$ is a continuous variable to express the incremental changes of shunt capacitor $k$ from time $t-1$ to $t$, and $z_{k,t}$ is a binary variable to denote if there exists incremental changes of shunt capacitor $k$ from time $t-1$ to $t$. Eq. (\ref{eq_con_sc_switch_1}) enforces $z_{k,t}$ to be one if $Q^{\text{SC}}_{k,t}$ and $Q^{\text{SC}}_{k,t-1}$ are different, where $M$ is a big positive number. Eqs. (\ref{eq_con_sc_switch_2})-(\ref{eq_con_sc_switch_5}) ensure that $\Delta^{\text{SC}}_{k,t}$ equals to $Q^{\text{SC}}_{k,t} - Q^{\text{SC}}_{k,t-1}$ if $z_{k,t}$ is one, and $\Delta^{\text{SC}}_{k,t}$ equals to zero when $z_{k,t}$ is zero. With the set of constraints, $\Delta^{\text{SC}}_{k,t}$ precisely denotes the incremental changes and can be minimized in the objective function.

The expert solver \texttt{Expert} takes the disturbance $d$ (the set of faulty lines  $\mathcal{E}_{\text{L,NS}}^{\text{F}}$), the initial tie-line status $\hat{u}^{\text{L}}_{l}$, where $\forall l\in\mathcal{E}_{\text{L,T}}$, and the step index $\mathcal{T}=[t_0, t_1, \cdots, T]$ as inputs and solver the following MIP problem
\begin{subequations}\label{eq_expert}
\begin{align}
&\max\sum_{t}\sum_{i}u^{\text{D}}_{i,t}P^{\text{D}}_{i}\\
&\text{subject to }(\ref{eq_con_distflow_1})-(\ref{eq_con_tieline_3})\quad\forall t\in\mathcal{T}\\
& u^{\text{SC}}_{k,t}=0\quad\forall k,\forall t\in\mathcal{T}\label{eq_expert_3}
\end{align}
\end{subequations}
where (\ref{eq_expert_3}) deactivate shunt capacitors since they will not be considered in \texttt{Expert}. The solution will provide a series of tie-line status $u^{\text{L}}_{l,t_0},u^{\text{L}}_{l,t_1},\cdots,u^{\text{L}}_{l,T}$ for $l\in\mathcal{E}_{\text{L,T}}$. Then, the optimal tie-line operating actions can be parsed as $a^{\text{L}}_{t_1},\cdots,a^{\text{L}}_{T}$. The \texttt{Env.Reset} function computes the system initial condition given a random generated faulty line set $\mathcal{E}_{\text{L,NS}}^{\text{F}}$
\begin{subequations}\label{eq_env_reset}
\begin{align}
&\max\sum_{t}\sum_{i}u^{\text{D}}_{i,t}P^{\text{D}}_{i}\\
&\text{subject to }(\ref{eq_con_distflow_1})-(\ref{eq_con_line})\quad\forall t\in[t_0]\\
& u^{\text{L}}_{l,t_0}=0\quad\forall l\in\mathcal{E}_{\text{L,T}}\label{eq_env_reset_3}\\
& u^{\text{SC}}_{k,t_0}=0 \quad\forall k
\end{align}
\end{subequations}
where Eq. (\ref{eq_env_reset_3}) ensures no tie-line actions under this initial stage. The \texttt{Env.Step} aims to restore the maximal load given the disturbance, a tie-line status and the load status from the previous step by solving the following problem
\begin{subequations}\label{eq_env_step}
\begin{align}
&\max\sum_{t}\sum_{i}u^{\text{D}}_{i,t}P^{\text{D}}_{i}\\
&\text{subject to }(\ref{eq_con_distflow_1})-(\ref{eq_con_line}), (\ref{eq_con_tieline_3})\quad\forall t\in[t_\tau]\\
& u_{t_\tau}^{\text{D}}\geq\hat{u}^{\text{D}}_{t_{\tau-1}}\label{eq_env_step_3}\\
&u^{\text{SC}}_{k,t_\tau}=0 \quad\forall k,\forall t
\end{align}
\end{subequations}
where $\hat{u}^{\text{D}}_{t_{\tau-1}}$ is the load status from the previous step, and Eq. (\ref{eq_env_step_3}) ensures the restored load will not be disconnected again.

Similarly, hybrid-action expert solver \texttt{hExp} solves the following MIP
\begin{subequations}\label{eq_h_expert}
\begin{align}
&\max\sum_{t}(\sum_{i}u^{\text{D}}_{i,t}P^{\text{D}}_{i} + w\sum_{k}\Delta_{k,t}^{\text{SC}})\\
&\text{subject to }(\ref{eq_con_distflow_1})-(\ref{eq_con_sc_switch})\quad\forall t\in\mathcal{T}
\end{align}
\end{subequations}
where $w$ is the weighting factor. The hybrid-action DSR environment \texttt{hEnv} also consider the reactive power dispatch. The \texttt{hEnv.Reset} function computes the system initial condition given a random generated faulty line set $\mathcal{E}_{\text{L,NS}}^{\text{F}}$
\begin{subequations}\label{eq_h_env_reset}
	\begin{align}
	&\max\sum_{t}\sum_{i}u^{\text{D}}_{i,t}P^{\text{D}}_{i}\\
	&\text{subject to }(\ref{eq_con_distflow_1})-(\ref{eq_con_line})\quad\forall t\in[t_0]\\
	& u^{\text{L}}_{l,t_0}=0\quad\forall l\in\mathcal{E}_{\text{L,T}}\label{eqh_env_reset_3}\\
	& Q^{\text{SC}}_{k,t_0}=0 \quad\forall k\label{eqh_env_reset_4}
	\end{align}
\end{subequations}
where Eqs. (\ref{eqh_env_reset_3}) and (\ref{eqh_env_reset_4}) ensure no restorative actions under this initial stage. The \texttt{hEnv.Step} aims to restore the maximal load given the disturbance, a tie-line status and the load status from the previous step by solving the following problem
\begin{subequations}\label{eq_h_env_step}
	\begin{align}
	&\max\sum_{t}\sum_{i}u^{\text{D}}_{i,t}P^{\text{D}}_{i}+\sum_{t}\sum_{k}|e^{\text{SC}}_{k,t_\tau}|\\
	&\text{subject to }(\ref{eq_con_distflow_1})-(\ref{eq_con_line}), (\ref{eq_con_tieline_3})\quad\forall t\in[t_\tau]\\
	& u_{t_\tau}^{\text{D}}\geq\hat{u}^{\text{D}}_{t_{\tau-1}}\\
	& e^{\text{SC}}_{k,t_\tau} = Q^{\text{SC}}_{k,t_\tau}-\hat{Q}^{\text{SC}} \quad\forall k
	\end{align}
\end{subequations}
where $\hat{u}^{\text{D}}_{t_{\tau-1}}$ is the load status from the previous step, and $\hat{Q}^{\text{SC}}$ is the var dispatch command. To avoid dispatch infeasibility due to the unenergized islands, the absolute error between the expert signal $\hat{Q}^{\text{SC}}$ and the actual var dispatch $Q^{\text{SC}}_{k,t_\tau}$ is minimized. Equivalent formulations to remove the absolute operator are implemented.

\section{Case Study}\label{sec_case}
In the numerical experiments, two metrics are considered to evaluate the learning performance: (1) \emph{Restoration ratio}: the ratio between the restored load by the agent and the optimal restorable load by the expert; (2) \emph{Restoration value}: total restored load by the agent in each episode; (3) \emph{Success rate}: number of times that the agent achieves optimal restorable load in $N$ episodes. The optimization is formulated using Pyomo \cite{hart2017pyomo} (National Technology and Engineering Solutions of Sandia, LLC, U.S.) and solved using IBM ILOG CPLEX 12.8. The deep learning model is built using TensorFlow $r1.14$.

It is worth mentioning that the scalability of the proposed method is demonstrated from both disturbance complexity and system size perspectives. The disturbance complexity determines the number of randomly tripped lines in each episode. The agent will encounter larger numbers of different topologies if more lines are randomly tripped. The system size verifies if the method can handle considerable sizes of inputs and larger action space.

\subsection{33-Bus System}
\begin{figure}[h]
	\centering
	\includegraphics[scale=0.2]{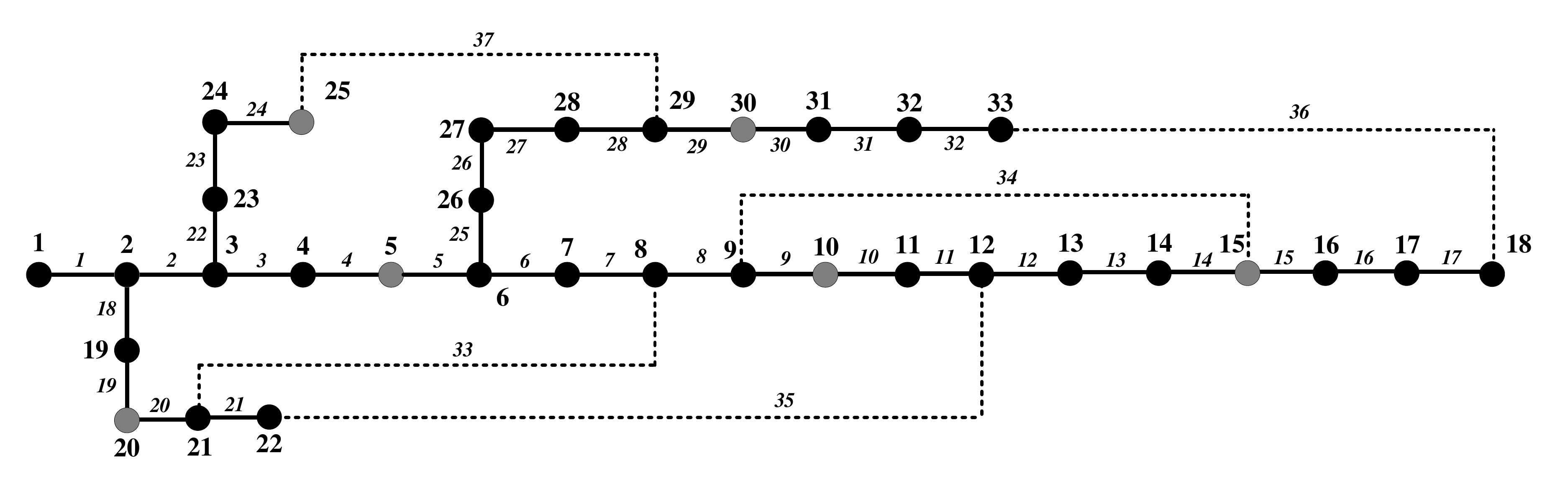}
	\caption{The 33-node system with five tie-lines. The gray nodes are assumed to be equipped with shunt capacitors.}
	\label{fig_33Feeder}
\end{figure}

The 33-bus system in \cite{Stentz1988} will be employed for the first case study. It is a radial 12.66 kV distribution network, shown in Fig. \ref{fig_33Feeder}. Detailed network data can be found in \cite{Stentz1988}. In this system, there are five tie-lines, which are assumed to be opened in the initial phase. Six shunt capacitors are assumed to be deployed in the gray nodes Fig. \ref{fig_33Feeder}. The dispatch ranges of all shunt capacitors are from -0.2 to 0.2 MVar. We assume the substation voltage is 1.05 p.u., and the voltage deviation limit is 0.05 p.u.
\subsubsection{Policy Network and Feature Selection}
\begin{figure}[h]
	\centering
	\includegraphics[scale=0.6]{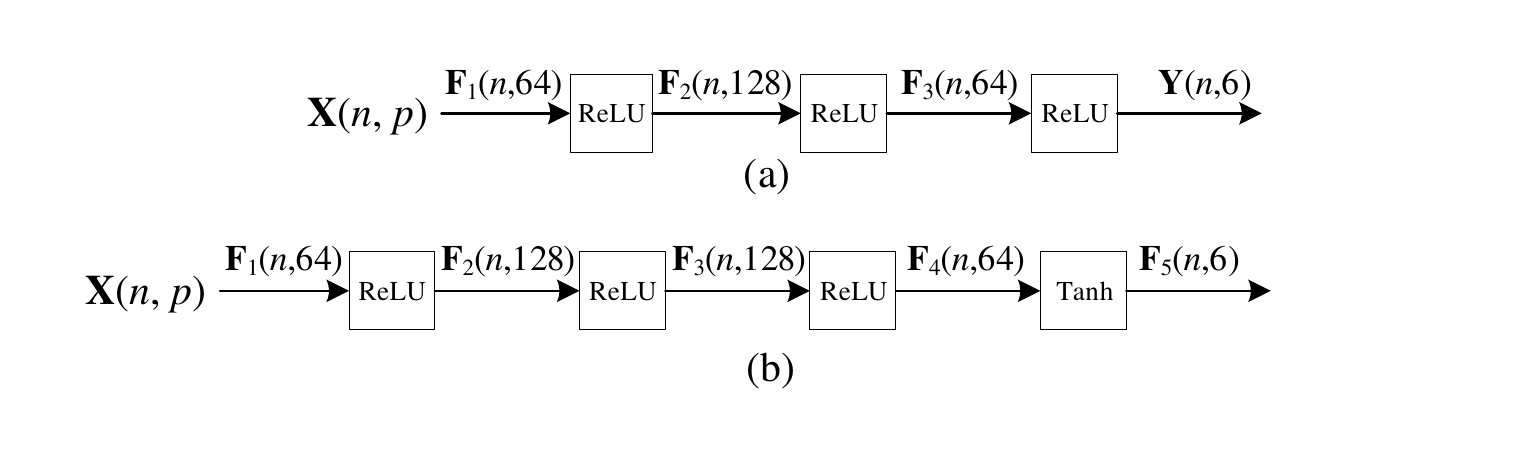}
	\caption{Deep neural network based policy networks. (a) Tie-line operation policy network. (b) Shunt capacitor dispatch policy network.}
	\label{fig_dnn}
\end{figure}

Based on the system structure, the policy networks are shown in Fig. \ref{fig_dnn}. The tie-line operation policy network consists of three hidden layers. We use the rectifier linear units (\texttt{relu}) as our activation functions. For the tie-line operation, the connectivity of the system is essential, and thus the feature inputs are line status. The shunt capacitor policy network has four hidden layers. For this network, load status and real-valued power flow are considered as feature inputs to extract the reactive power effects on the local voltage and load pick-up. Two types of activation functions are also compared.
\begin{figure}[h!]
	\centering
	\subfloat[]{
		\begin{minipage}[]{0.22\textwidth}
			\centering
			\includegraphics[scale=0.18]{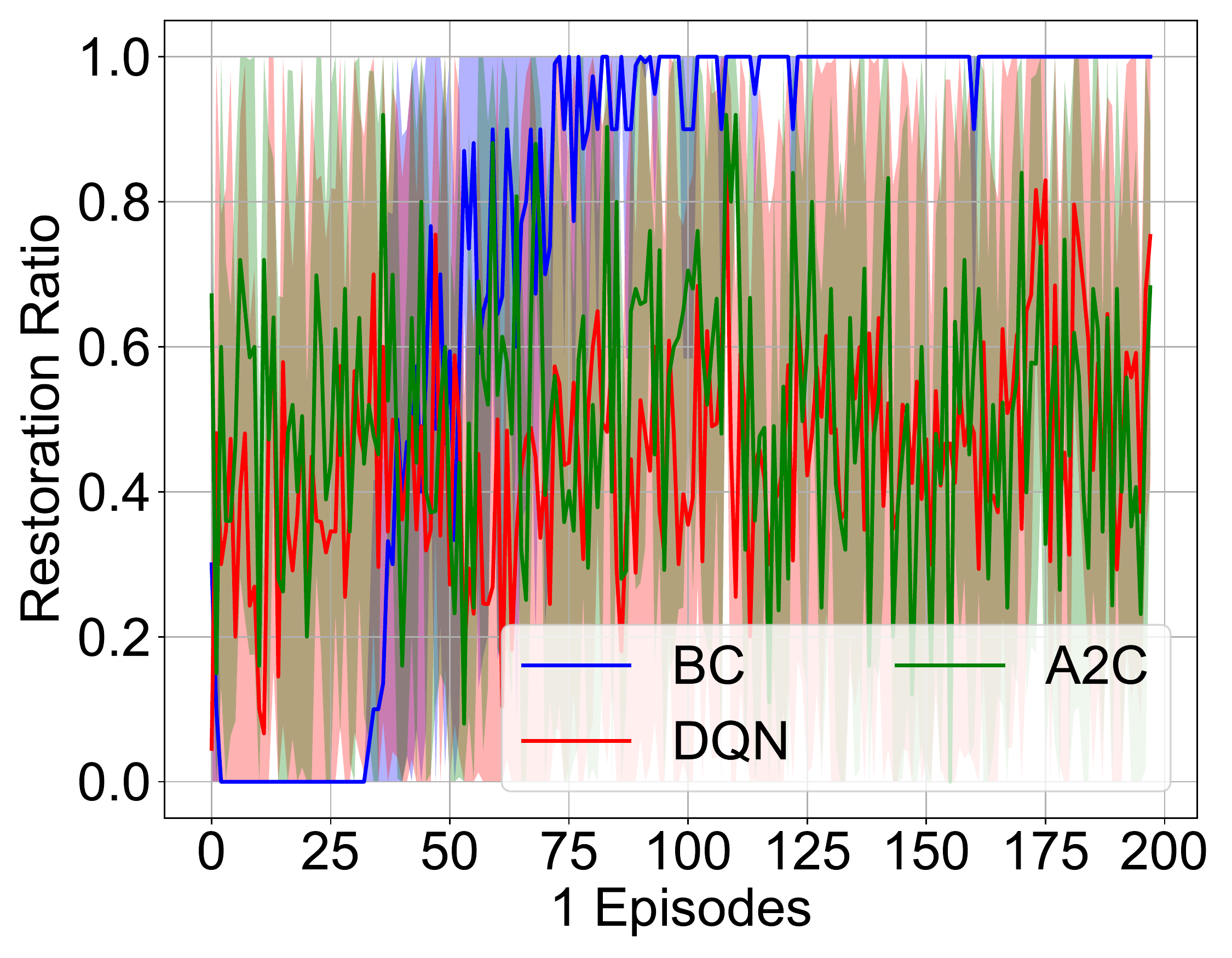}
		\end{minipage}
	}
	\subfloat[]{
		\begin{minipage}[]{0.22\textwidth}
			\centering
			\includegraphics[scale=0.18]{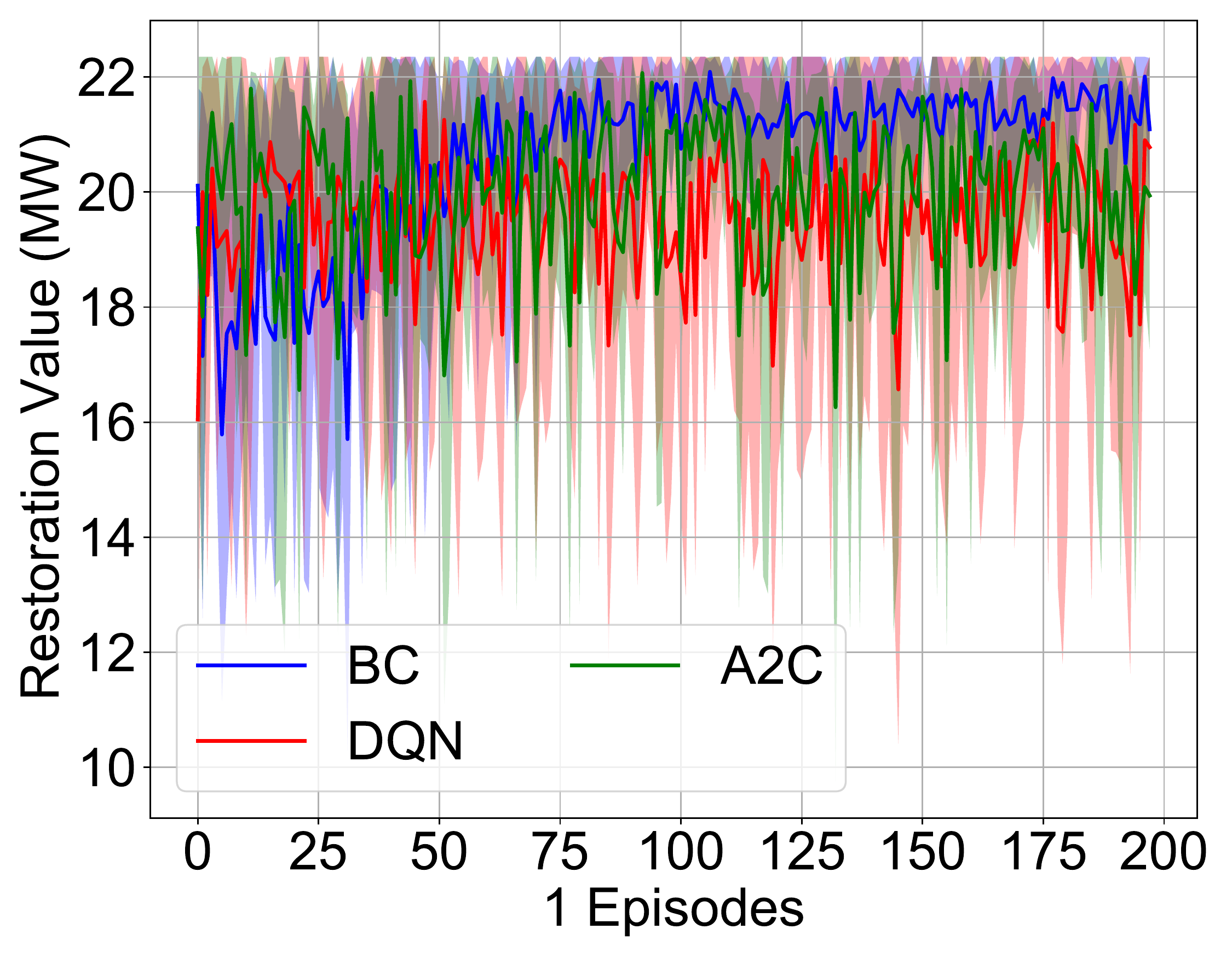}
		\end{minipage}
	}
	\caption{Training performance of imitation learning BC and reinforcement learning DQN and A2C under the $N-1$ scenario. (a) Restoration ratio. (b) Restoration value.}
	\label{fig_il_rl}
\end{figure}
\begin{figure}[h!]
	\centering
	\subfloat[]{
		\begin{minipage}[]{0.22\textwidth}
			\centering
			\includegraphics[scale=0.18]{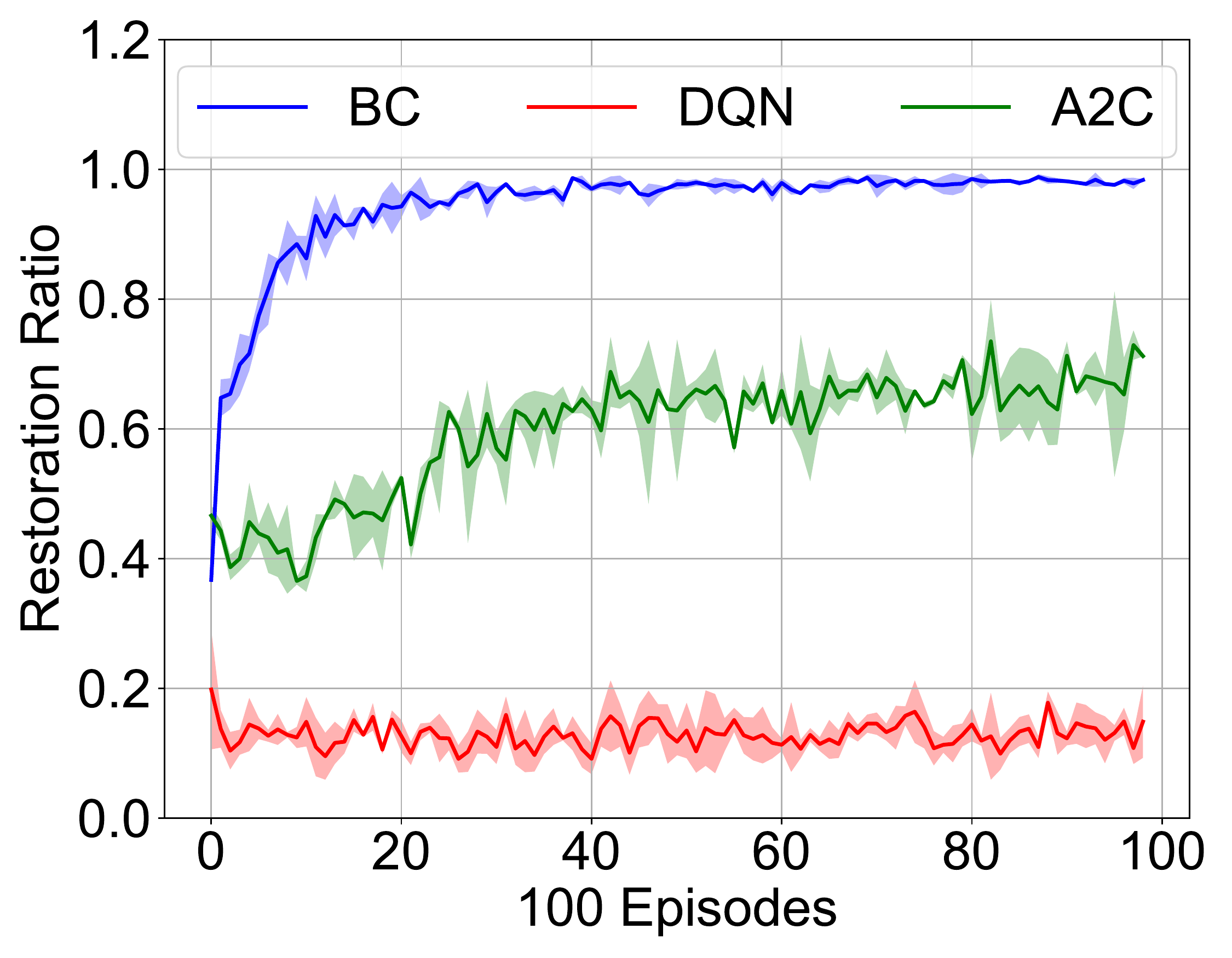}
		\end{minipage}
	}
	\subfloat[]{
		\begin{minipage}[]{0.22\textwidth}
			\centering
			\includegraphics[scale=0.18]{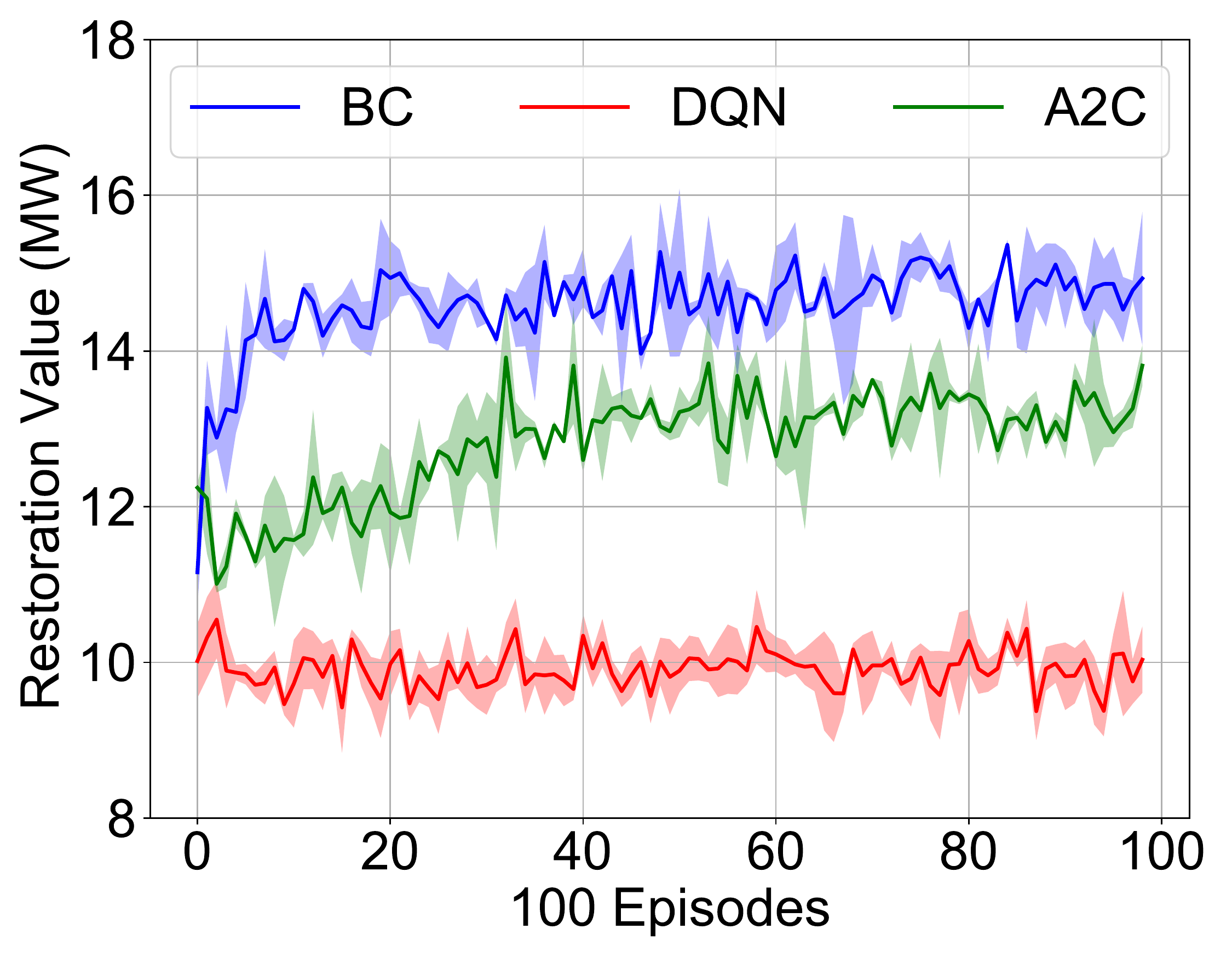}
		\end{minipage}
	}
	\caption{Training performance of imitation learning BC and reinforcement learning DQN and A2C under the $N-5$ scenario. (a) Restoration ratio. (b) Restoration value.}
	\label{fig_il_rl_n_5}
\end{figure}
\begin{figure}[h!]
	\centering
	\subfloat[]{
		\begin{minipage}[]{0.22\textwidth}
			\centering
			\includegraphics[scale=0.18]{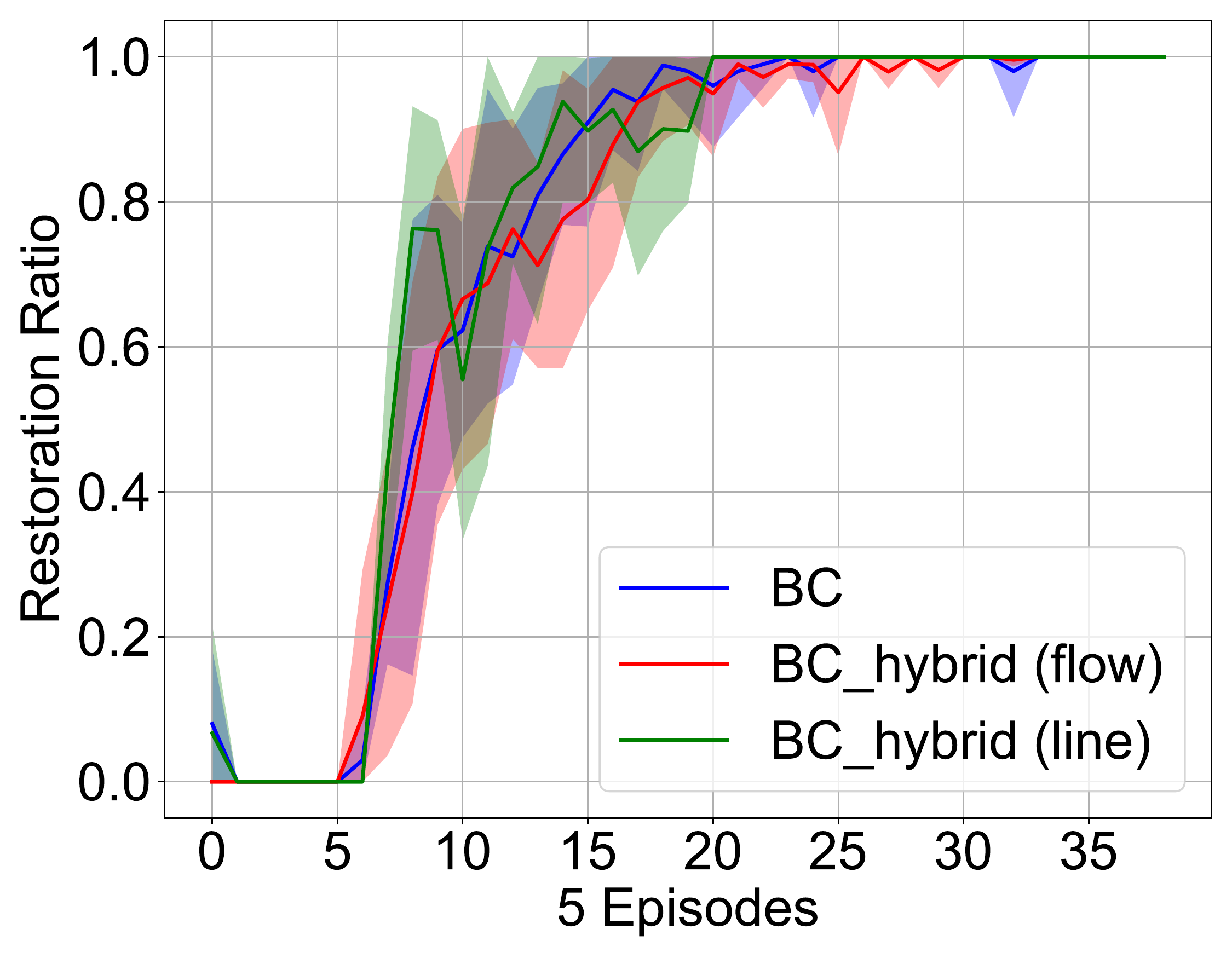}
		\end{minipage}
	}
	\subfloat[]{
		\begin{minipage}[]{0.22\textwidth}
			\centering
			\includegraphics[scale=0.18]{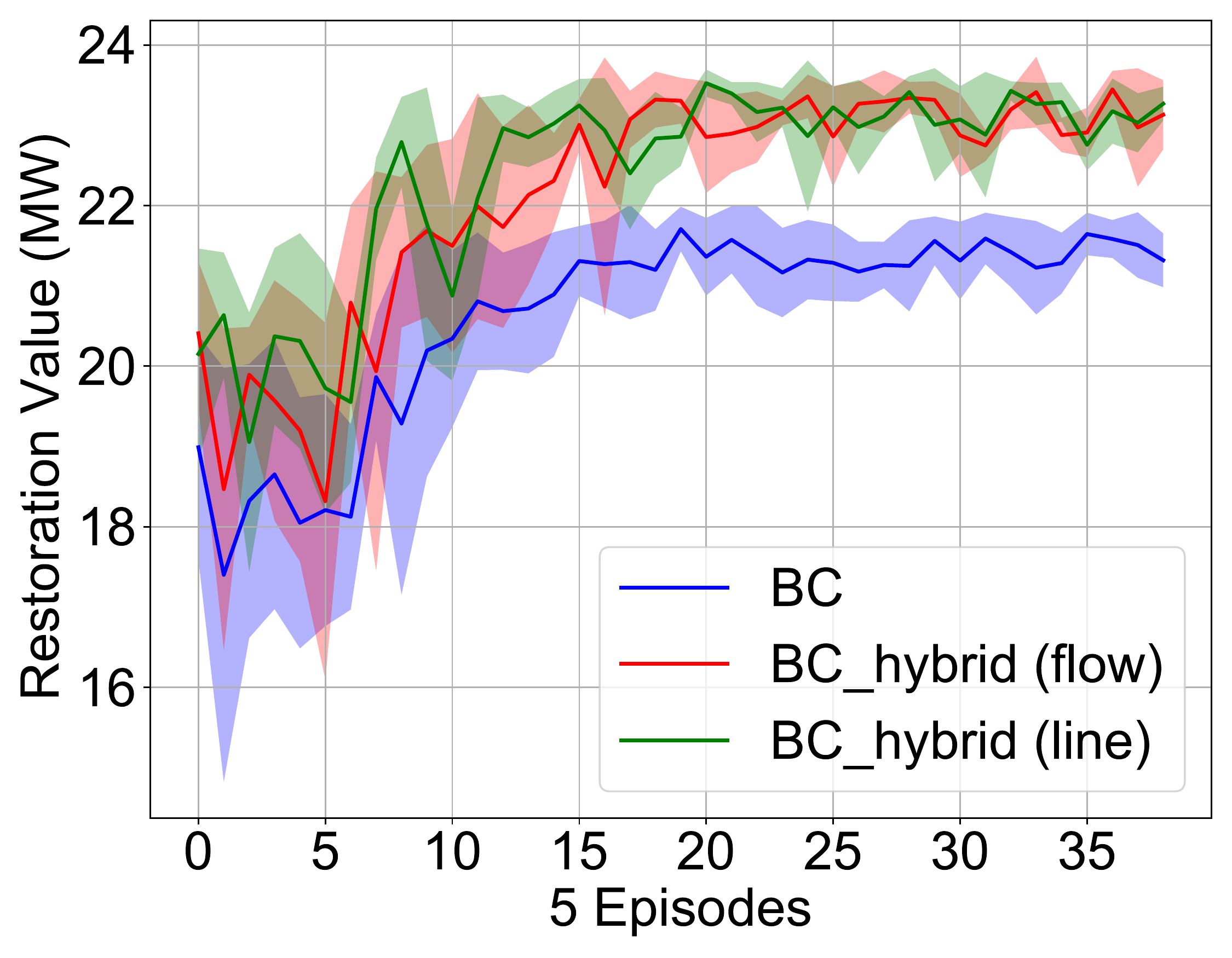}
		\end{minipage}
	}
	\caption{Training performance of BC and HBC under the $N-1$ scenario. (a) Restoration ratio. (b) Restoration value.}
	\label{fig_n_1}
\end{figure}

\begin{figure}[h!]
	\centering
	\subfloat[]{
		\begin{minipage}[]{0.22\textwidth}
			\centering
			\includegraphics[scale=0.18]{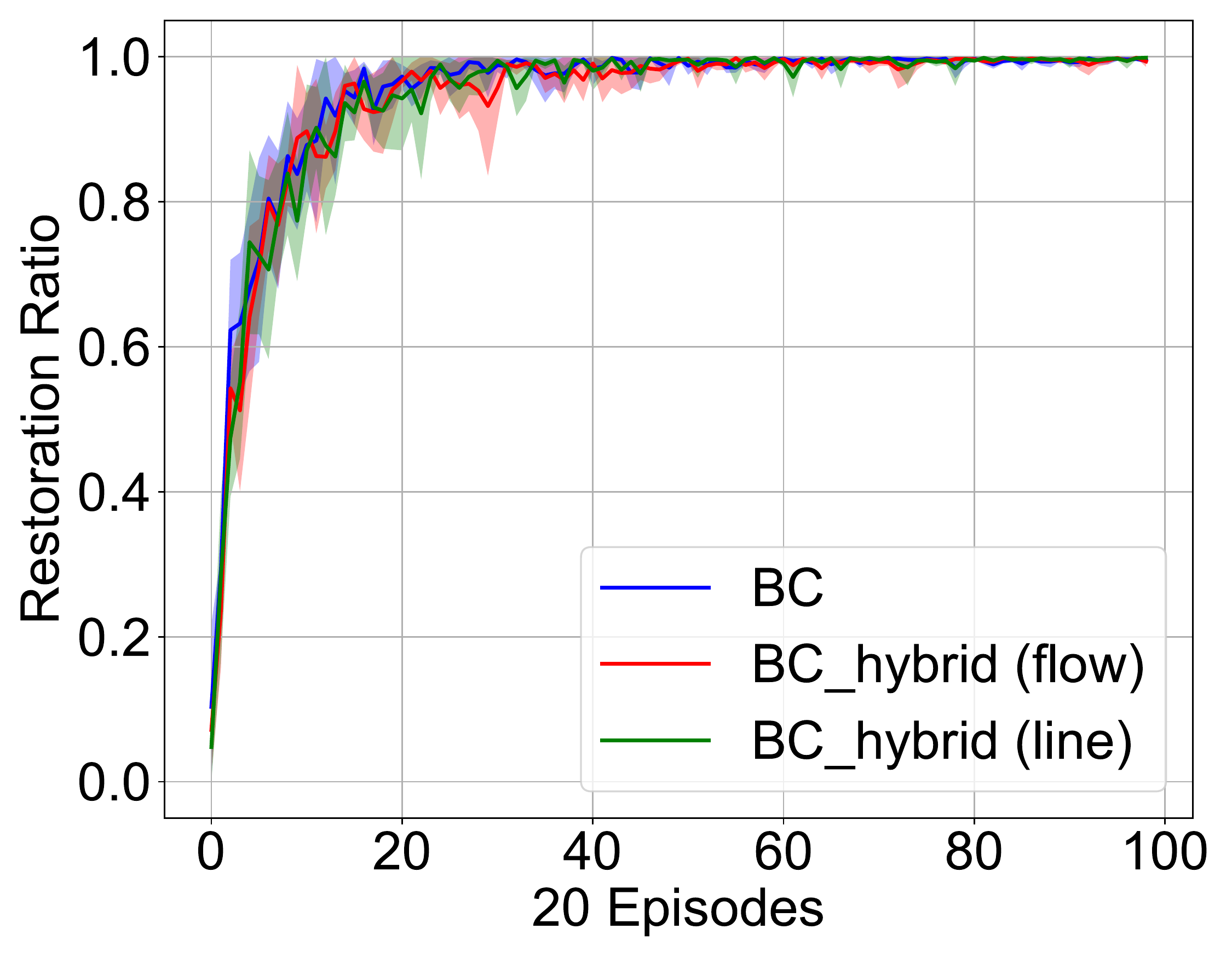}
		\end{minipage}
	}
	\subfloat[]{
		\begin{minipage}[]{0.22\textwidth}
			\centering
			\includegraphics[scale=0.18]{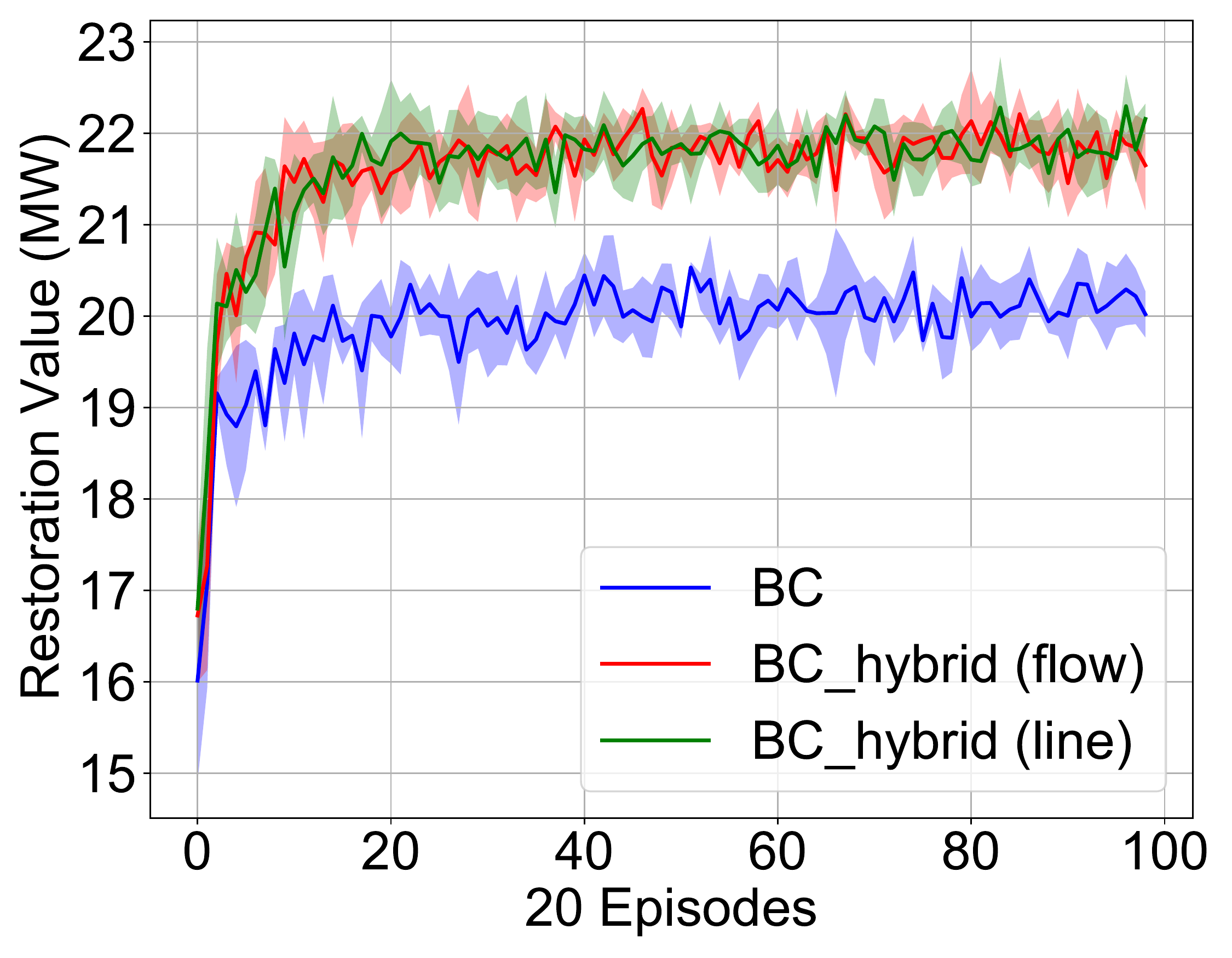}
		\end{minipage}
	}
	\caption{Training performance of BC and HBC under the $N-2$ scenario. (a) Restoration ratio. (b) Restoration value.}
	\label{fig_n_2}
\end{figure}

\subsubsection{IL v.s. RL for Network Reconfiguration under N-1 and N-5 Contingencies}
In this subsection, we compare the imitation learning Algorithms \ref{algo_bc} with two RL baseline algorithms, deep Q-network (DQN) and advanced actor critic (A2C). The DQN and A2C are implemented based on the reference in \cite{stable-baselines}. The $N-1$ random contingency is considered first. The total training episodes are 200. The restoration ratio and value are shown in Fig. \ref{fig_il_rl} (a) and (b), respectively. As shown, the BC algorithm is able to optimally restore the system after 75 episodes of training, while DQN and A2C admit only an averaged 45\% restoration ratio over the 200 episodes. The problem complexity due to the topology switching is intractable for algorithms that heavily rely on exploration like traditional RL.

For further verification, the $N-5$ random contingency is applied. Ten thousand training episodes are used. The restoration ratio and value are shown in Fig. \ref{fig_il_rl_n_5} (a) and (b), respectively. With increasing complexity of the problem, the advantage of BC compared with DQN is more significant as BC can achieve more than 90\% restoration ratio while DQN stays at 15\%. A2C performs better than DQN in the $N-5$ scenario and still underperforms the IL with a 35\% restoration ratio deficit.

\subsubsection{System Status during Restoration}
\begin{figure}[h]
	\centering
	\includegraphics[scale=0.25]{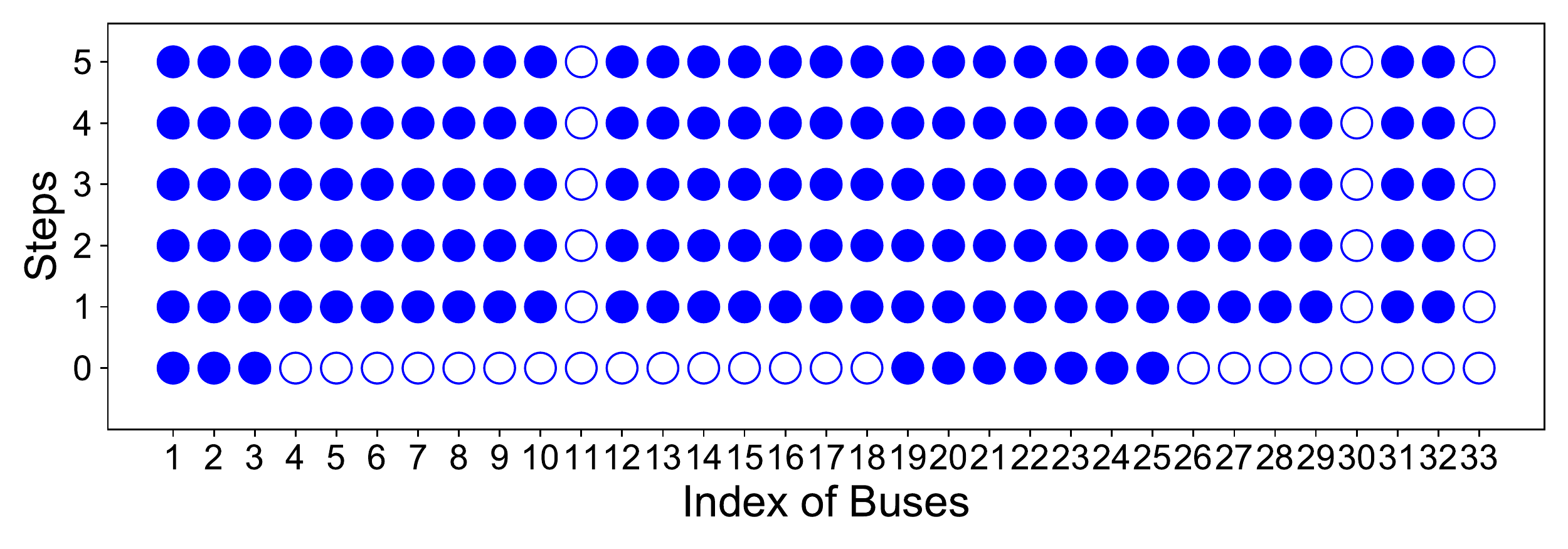}
	\caption{System load energization status under Line 3 outage.}
	\label{fig_op_n_1_load}
\end{figure}

\begin{figure}[h]
	\centering
	\includegraphics[scale=0.25]{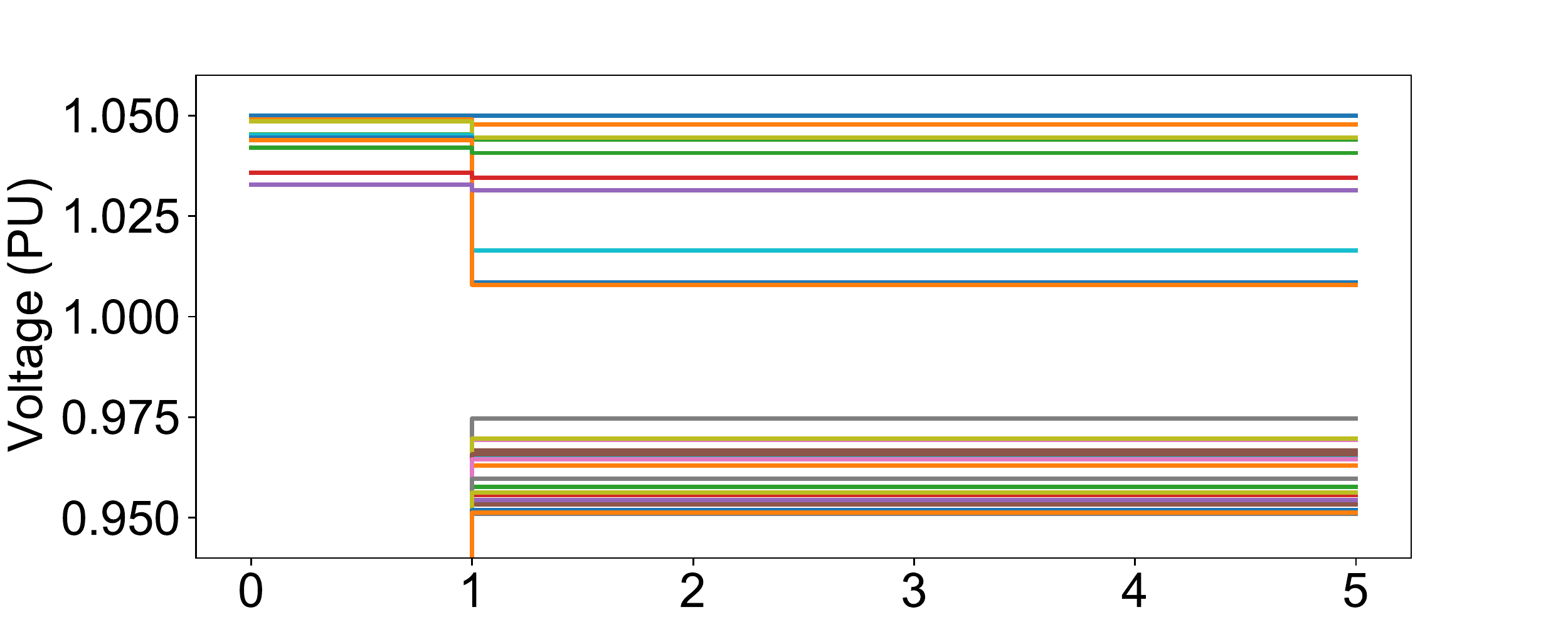}
	\caption{System voltage profile under Line 3 outage.}
	\label{fig_op_n_1_volt}
\end{figure}

A particular scenario of $N-1$ contingency is illustrated. In this scenario, Line 3 is damaged and tripped. Once the damage situation is transmitted, the agent closes Tie-line 33 in the first step and performs no further actions in the following steps to respect the radiality constraint. The corresponding load energization status and voltage profile are shown in Fig. \ref{fig_op_n_1_load} and \ref{fig_op_n_1_volt}, respectively. Since Line 3 is upstream of the network, its outage causes de-energization of 60\% load. Fortunately, with the reconfiguration through Tie-line 33, most of the loads have been picked up except for Loads 11, 30, and 33. It is because energizing these loads will violate the voltage security constraint as shown in Fig. \ref{fig_op_n_1_volt}, which indicates the necessity of voltage compensation.
\begin{figure}[h]
	\centering
	\includegraphics[scale=0.25]{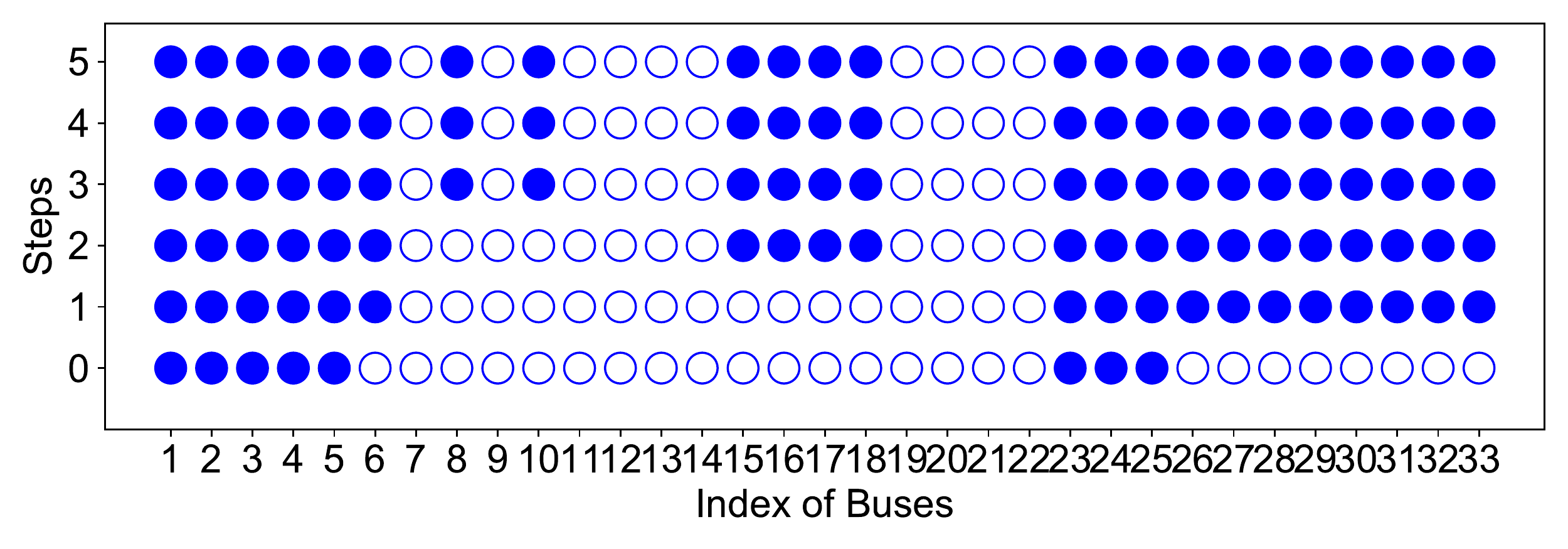}
	\caption{System load energization status under Lines 5, 6, 10, 14, and 18 outages.}
	\label{fig_op_n_5_load}
\end{figure}
\begin{figure}[h]
	\centering
	\includegraphics[scale=0.25]{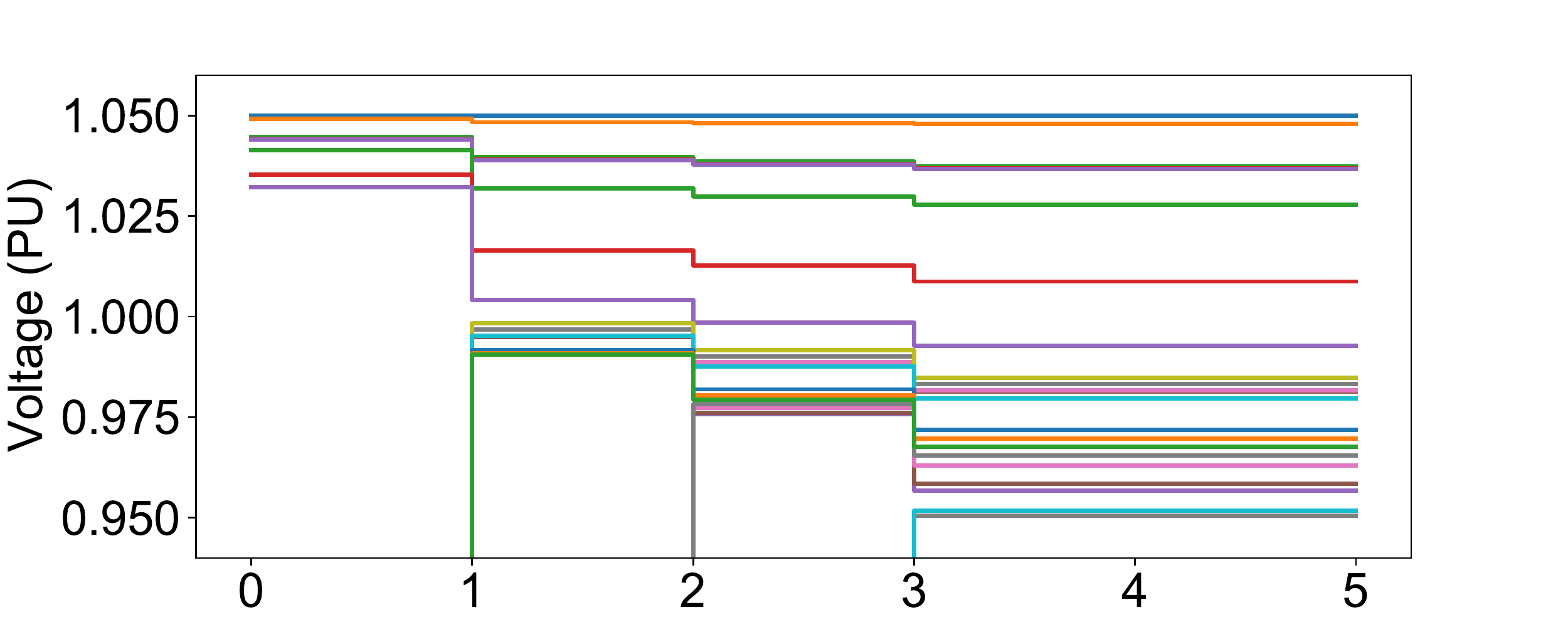}
	\caption{System voltage profile under Lines 5, 6, 10, 14, and 18 outages.}
	\label{fig_op_n_5_volt}
\end{figure}

Another scenario of $N-5$ contingency is described, where Lines 5, 6, 10, 14, and 18 are tripped. In response, the agent then closes Tie-lines 37, 36, 34, 35, and 33 in sequence. As shown in Fig. \ref{fig_op_n_5_load}, several loads can be energized after each step. At last, ten loads cannot be picked up due to the voltage constraint, although all buses are connected through the tie-lines. The voltage profile is depicted in Fig. \ref{fig_op_n_5_volt}.

\subsubsection{Hybrid Policy for N-1 and N-2 Contingencies}
Under the random $N-1$ contingency, the hybrid policy network is trained for 200 episodes. In the var dispatch policy network, two features are considered: load status and real-valued power flow. The training performance is illustrated in Fig. \ref{fig_n_1}. All metrics are averaged within five steps. The BC algorithm has a lower variation in the restoration ratio since the task only involves discrete actions and relatively easier. But with var dispatch capability, the hybrid agent is able to restore approximately 2 MW load in each episode as shown in Fig. \ref{fig_n_1} (b). As for the features, real-valued power flow and the load status have the similar performance.

A more complicated random $N-2$ scenario is considered and train both BC and HBC agents for 2000 episodes. Similarly, BC has a lower variation in the restoration ratio, particularly when all algorithms achieve high restoration ratio at around 400 episodes as shown in Fig. \ref{fig_n_2}.  Fig. \ref{fig_n_2} (b) shows that the HBC agent can restore 2 MW more in each episode, indicating that it is critical to have var support in the resilient setting. The reason lies in the fact that the reconfigured network may have longer feeders when there are more line outages. Therefore, the voltage drops along reconfigured feeders are more significant. 

\begin{figure}[h]
	\centering
	\includegraphics[scale=0.3]{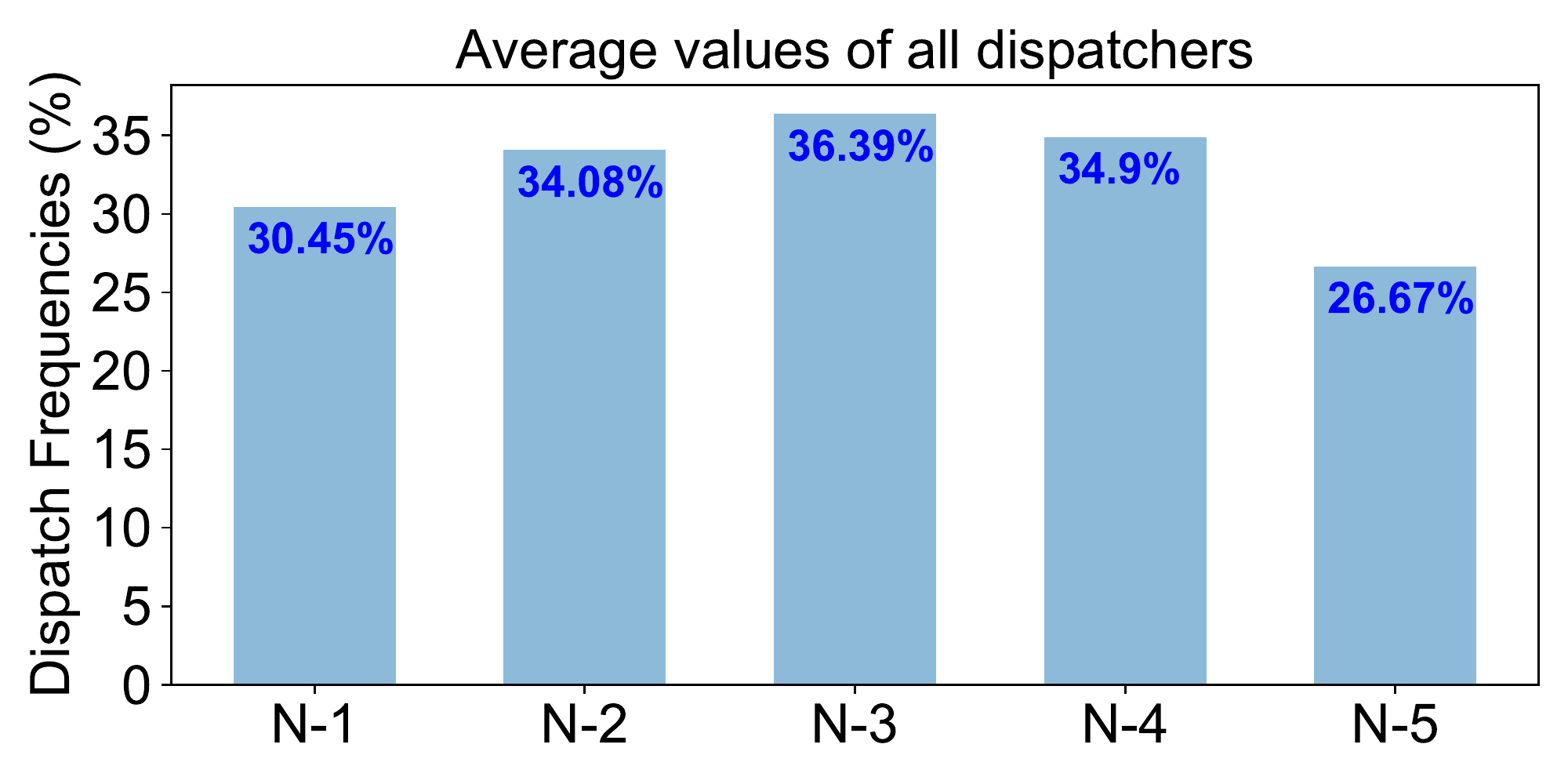}
	\caption{Averaged var dispatch frequencies of all shunt capacitor dispatchers.}
	\label{fig_7_1_r2}
\end{figure}

It is worth noting that with increasing numbers of line damages the usage of shunt capacitors can be reduced since the designated shunt capacitors may result in one of the unenergized islands. Frequencies of the reactive power control from $N-1$ to $N-5$ are calculated as in Fig. \ref{fig_7_1_r2}. With increasing number of line outages, the usage of var dispatch first increases as longer feeder may occur, and then decreases due to the unenergized islands.

\subsection{119-Bus System}
\begin{figure}[h!]
	\centering
	\subfloat[]{
		\begin{minipage}[]{0.22\textwidth}
			\centering
			\includegraphics[scale=0.18]{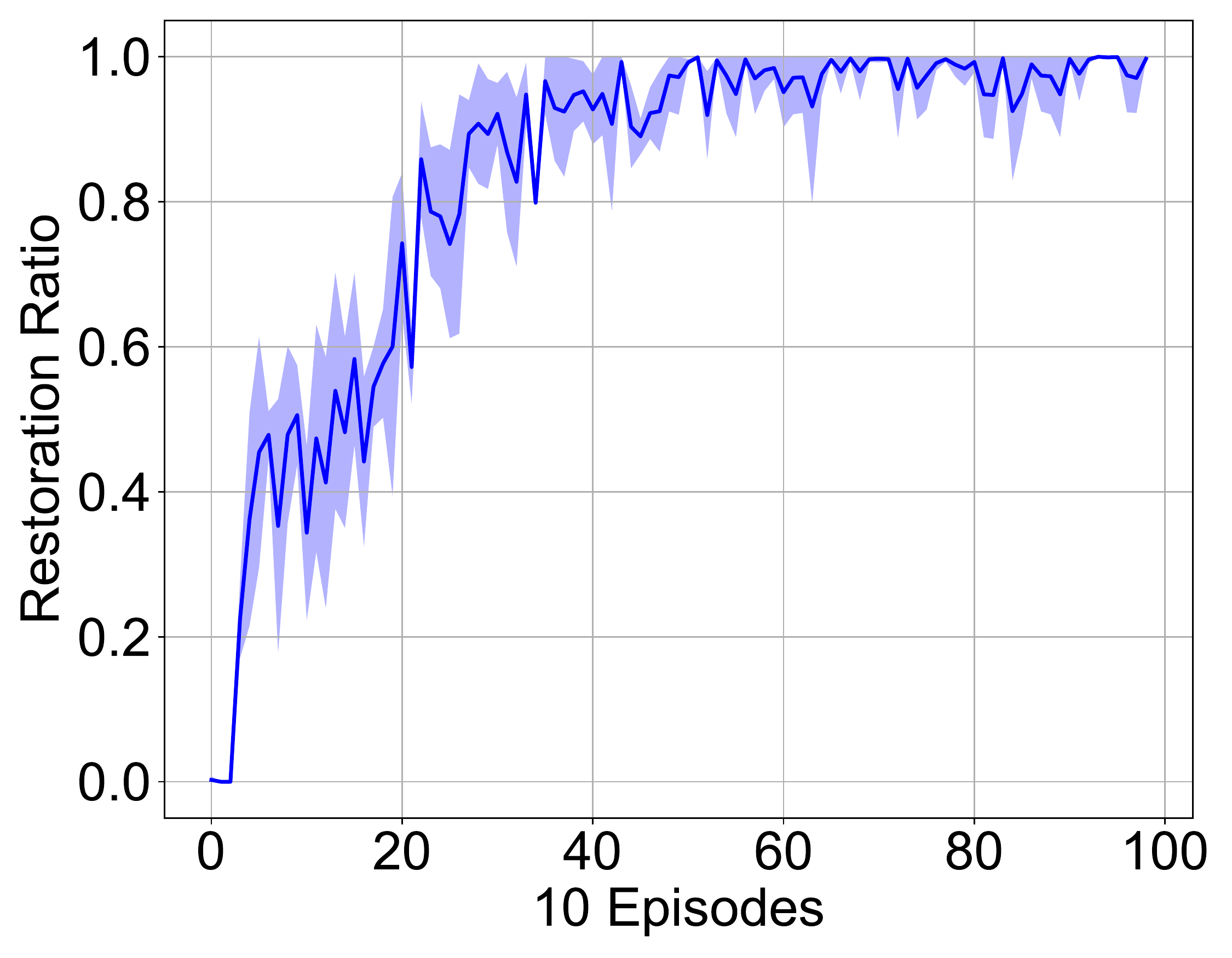}
		\end{minipage}
	}
	\subfloat[]{
		\begin{minipage}[]{0.22\textwidth}
			\centering
			\includegraphics[scale=0.18]{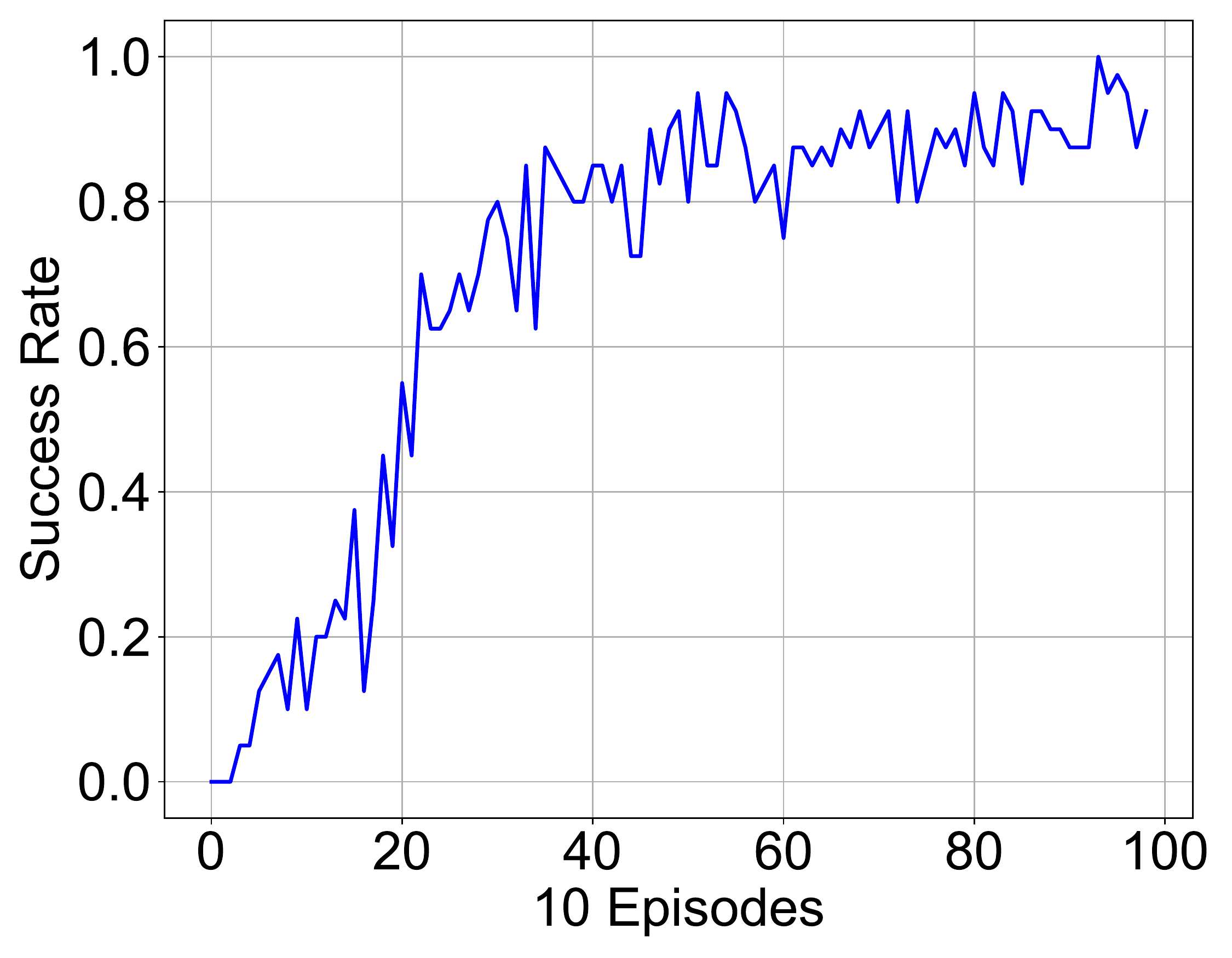}
		\end{minipage}
	}
	\caption{Training performance of BC under the $N-1$ scenario. (a) Restoration ratio. (b) Success rate.}
	\label{fig_IL_119bus}
\end{figure}

The second case study is demonstrated on the 119-bus system. This system is an 11 kV distribution system with 15 tie-lines, which is particularly suitable to study network reconfiguration and restoration. Detailed data of the system can be found in \cite{Zhang2007}. We assume the substation voltage is 1.05 p.u., and the voltage deviation limit is 0.05 p.u.
The $N-1$ random line outage is considered, and the tie-line operation is considered for restoration. The restoration agent is trained for 1000 episodes. The results are illustrated in Fig. \ref{fig_IL_119bus}. The agent is able to achieve an 80\% ratio after 400 episodes and an 80\% success rate after 600 episodes. Additionally, the effect of integer numbers in MP-based expert on the training performance is studied. The results of the 33-bus are used for comparison.  The episode after which the success rate is above 80\% is considered as a metric, denoted as the \emph{confident episode}. The results are summarized below in Table \ref{tab_integer_sensitivity}.  The confident episode of 33-bus is 75, while the one of 119-bus is 600. The ratio of the confident episode between 33-bus and 119-bus, which is eight, is close to the one of tie-line integer numbers between these two cases, which is nine. Intuitively, the integer variables that control the tie-lines should have a dominant impact on the IL performance. 
\begin{table}
	\caption{Effect of Integer Numbers in MP-based Expert on IL Performance}
	\centering
	\begin{tabular}{lclclclcl}
		\toprule 
		Case & Integer number & Tie-line integer number  & Confident episode  \\
		\midrule
		33-bus & 6960 & 25 & 75 \\
		119-bus & 226800 & 225 & 600 \\
		\bottomrule
	\end{tabular}
	\label{tab_integer_sensitivity}
\end{table}

\section{Conclusions and Future Works}\label{sec_con}
In this paper, we propose the IL framework and HBC algorithm for training intelligent agents to perform online service restoration. We strategically design the MP-based experts, who are able to provide optimal restoration actions for the agent to imitate, and a series of MP-based environments that agents can interact with. Agents that are trained under the proposed framework can master the restoration skills faster and better compared with RL methods. The agent can perform optimal tie-line operations to reconfigure the network and simultaneously dispatch reactive power of shunt capacitors using the trained policy network. The decision-making process has negligible computation costs and can be readily deployed for online applications. Future efforts will be devoted to feature extraction capability considering unique power network structure as well as a multi-agent training paradigm to incorporate distributed energy resources for the build-up restoration strategy.

\bibliography{IEEEabrv_zyc,library,Ref_ILRL}
\bibliographystyle{IEEEtran}

\end{document}